\documentclass[prd,aps,twocolumn,floats,floatfix,nofootinbib,superscriptaddress]{revtex4-2}

\usepackage[english]{babel}

\usepackage{amsmath}
\usepackage{graphicx}
\usepackage{amssymb}
\usepackage{verbatim}
\usepackage[colorlinks=true, allcolors=blue]{hyperref}
\usepackage[utf8]{inputenc}
\usepackage{float}
\usepackage{ulem}
\usepackage{cancel}
 
\def\Journal#1#2#3#4{{#1} {\bf #2}, #3 (#4)}

\def\LRR{Liv. Rev. Relat.}

\def\NPB{{Nucl. Phys.} B}
\def\PLB{{Phys. Lett.}  B}
\def\PLA{{Phys. Lett.}  A}
\def\PRL{Phys. Rev. Lett.}
\def\PRD{{Phys. Rev.} D}
\def\PR{{Phys. Rev.}}
\def\PRep{{Phys. Rep.}}

\def\APJ{{Astrophys. J.}}
\def\APJL{{Astrophys. J. Lett.}}
\def\AA{{Astron. Astrophys.}}
\def\MNRAS{{Mon. Not. R. Astr. Soc.}}
\def\NAT{{Nature}}
\def\SCI{{Science}}
\def\CQG{{Class. Quant. Grav.}}
\def\CMP{{Commun. Math. Phys.}}
\def\JCAP{{JCAP}}
\def\JMP{{J. Math. Phys.}}

\def\IJMPA{{Int. J. Mod. Phys.} A}

\def\l{\left}
\def\r{\right}

\newcommand{\be}{\begin{equation}}
\newcommand{\ee}{\end{equation}}
\newcommand{\bea}{\begin{eqnarray}}
\newcommand{\eea}{\end{eqnarray}}

\begin{document}

\title{\textbf{Nonlinear Stability of Rotating Hairy Black Holes}}

\author{Juan A. Carretero}
\email{juan-antonio.carretero@uib.cat}
\address{Departament  de  F\'{\i}sica,  Universitat  de  les  Illes  Balears,  Palma  de  Mallorca,  Baleares  E-07122,  Spain}
\address{Institute of Applied Computing \& Community Code (IAC3),  Universitat  de  les  Illes  Balears,  Palma  de  Mallorca,  Baleares  E-07122,  Spain}

\author{Philippe Grandclément}
\email{Philippe.Grandclement@obspm.fr}
\address{LUX, Observatoire de Paris, Université PSL, Sorbonne Université, CNRS, 92190 Meudon, France}

\author{Carlos Palenzuela}
\email{Corresponding author: carlos.palenzuela@uib.es}
\address{Departament  de  F\'{\i}sica,  Universitat  de  les  Illes  Balears,  Palma  de  Mallorca,  Baleares  E-07122,  Spain}
\address{Institute of Applied Computing \& Community Code (IAC3),  Universitat  de  les  Illes  Balears,  Palma  de  Mallorca,  Baleares  E-07122,  Spain}

\author{Marcelo Salgado}
\email{marcelo@nucleares.unam.mx}
\address{Instituto de Ciencias Nucleares, Universidad Nacional Aut\'onoma de M\'exico, Circuito Exterior C.U., A.P. 70-543, M\'exico D.F. 04510, M\'exico.}
\address{Departament  de  F\'{\i}sica,  Universitat  de  les  Illes  Balears,  Palma  de  Mallorca,  Baleares  E-07122,  Spain}

\begin{abstract}
Rotating hairy black holes (RHBHs) are axisymmetric equilibrium solutions of the Einstein–Klein–Gordon equations, consisting of a spinning black hole surrounded by a toroidal distribution of complex scalar field. Despite their potential astrophysical relevance, the stability of these configurations—naturally expected to form through superradiant growth of light bosonic fields—remains uncertain.
In this work, we investigate the stability of RHBHs  by performing fully non-linear numerical evolutions of several configurations that differ in the relative mass contribution of the scalar-field torus. We find that configurations in which the scalar field mass is subdominant compared to the black hole mass remain stable throughout the evolution, at least on timescales of order $\mu t > 1600$, where $\mu$ is the scalar field mass. These configurations might therefore be stable, with possible superradiant instabilities developing only on much longer timescales $\mu t \sim 10^{11}$, according to previous linear stability analyses. In contrast, when the scalar-field mass dominates, the system develops an instability in a much shorter timescale around $\mu t \sim {\cal O} (100)$,
similar to the non-axisymmetric instability observed in rotating boson stars.
Given the expected upper limits on scalar-field mass growth achievable through superradiance, our results suggest that rotating hairy black holes formed predominantly via this mechanism are likely to remain stable, at least up to the onset of the superradiant instability.
\end{abstract}

\maketitle

\section{Introduction}
Black holes (BH) and gravitational waves (GW) are two of the most striking predictions of Einstein's general theory of relativity (GR). Remarkably, these two 
seemingly unrelated concepts are intertwined together in view that one of the most powerful sources of gravitational radiation is precisely a binary BH system: after spiraling and losing energy due to a continuous emission of GW, the two BHs finally collide and merge into a single BH, which is  accompanied by an additional burst of gravitational radiation. This kind of events have been corroborated observationally in multiple occasions through the direct detection of GW by the LIGO-Virgo-KAGRA (LVK) collaboration, from the very first detection in 2015 to the latest one ten years after in 2025 \cite{VLK}. The analysis of observational data, together with numerical simulations solving the full nonlinear Einstein field equations \cite{Ninja1}, has been complemented by approximate approaches such as post-Newtonian \cite{PPN} and effective-one-body methods \cite{EOB}. These combined efforts have led to the conclusion that the detected gravitational-wave signals originate primarily from binary BHs, and in some cases from binary compact objects, some of which are likely neutron stars \cite{BSNS}. In the binary BH case, the BH masses range 
from a few solar masses to hundred solar masses each.
There exists also robust independent evidence about the existence of BHs in the universe from the observation of {\it shadows} around the supermassive object at the centers of some galaxies which are produced by the strong bending of light emitted from the accretion disk around the central object \cite{EHT}. 
The BH scenario at the center of galaxies, together with some additional ingredients, is also the simplest mechanism to explain the emission of relativistic jets in this kind of environments \cite{EHTjets}. Another evidence 
comes from the dynamics of stars that orbit around the center of our galaxy \cite{Ghezetal,Genzeletal}. 

All these observational confirmations are also consistent with the Kerr hypothesis --namely, that the initial binary black hole system consists of two spinning Kerr BHs, some of which may be slowly rotating and therefore close to the Schwarzschild limit. These BHs ultimately collide and merge into a single Kerr BH. The GW emitted by the remnant BH before it settles into a stationary state are consistent with linear perturbation theory and correspond to the so-called black-hole quasinormal modes (QNMs) of oscillation \cite{Berti2025}. These QNMs serve as a distinctive fingerprint of the original binary black-hole system. This broad framework of observations and theoretical results is consistent with the well-known no-hair conjecture \cite{Carter1968}, which states that stationary, asymptotically flat (AF), and rotating black hole solutions in general relativity belong to the Kerr-Newman family, characterized solely by three parameters: mass, angular momentum, and electric charge. Since most astrophysical objects are expected to discharge rapidly, the only physically relevant solutions are the Kerr black hole and its non-rotating limit, the Schwarzschild black hole.

This conjecture is supported by the uniqueness theorems \cite{uniqueness,Heusler1996} in GR and also by the no-hair theorems \cite{nohairtheorems,Pena1997}. Roughly speaking, the latter establish that, under fairly general and reasonable physical assumptions, certain additional fields around a BH cannot coexist with it in a self-consistent manner. Consequently, such fields must vanish identically, implying that the only possible black hole solutions in general relativity are vacuum ones (i.e., the Kerr or Schwarzschild solutions). On the other hand, 
the uniqueness theorems corroborate the no-hair conjecture associated with the existence of 
axisymmetric rotating AFBH solutions (i.e. the Kerr-Newman family) in the Einstein-electrovacuum system. Nevertheless, recent theoretical developments have shown that some of these no-hair theorems can be circumvented when certain underlying assumptions are relaxed \cite{Nucamendi2003}. 

Perhaps the most simple scenario in this direction, without introducing violations of the energy conditions, changing the asymptotic structure or even appealing to modifications of gravity, is by breaking some of the symmetry assumptions of the underlying spacetime. Notably, by introducing rotating scalar fields. Indeed, rotation is sufficient to evade the no-hair theorems in the Einstein-Klein-Gordon (EKG) system with complex valued scalar fields \cite{Pena1997}. The first rotating hairy BH (RHBH) solution of this kind was found numerically by Herdeiro \& Radu 
in 2014 \cite{Herdeiro2014,Herdeiro2015}. A simplified version of these solutions were also found when the background spacetime is fixed. This is the test-field approximation (sometimes called the decoupling limit), where the KG equation is solved 
in the exterior of a Kerr BH background. The resulting nontrivial solutions for the scalar field within this test-field approximation are termed cloud solutions \cite{Herdeiro2014,Herdeiro2015,Garcia2019,Garcia2020,Hod}. The cloud solutions can be extended to the electrically charged 
scenario in the background of a Kerr-Newman spacetime \cite{Benone2014,Garcia2023b}. Interestingly, the EKG system, when considered in the absence of an event horizon but under the same symmetry assumptions (e.g., stationarity and axisymmetry), gives rise to rotating boson star solutions (see Refs.~\cite{bosonstars} for a review). These are hypothetical, globally regular, and extended compact objects that have been proposed as potential candidates for part of the dark matter content of the Universe and as possible black hole mimickers \cite{mimickers}. Like the boson star, both the RHBH and the cloud configurations are ``quantized", in the sense that their solutions are labeled by the quantum numbers $(n,l,m)$. Due to this feature, analogous to the Hydrogen atom, this kind of systems are often referred to as {\it gravitational atoms}.

Thus, from the purely theoretical perspective, these novel hairy BH solutions within GR serve as counterexamples to the no-hair conjecture, demonstrating that it does not hold universally. However, this conclusion is still premature in several respects. First, from the physical point of view, the mere existence of such BH solutions is not sufficient to invalidate the conjecture; it is also necessary to demonstrate that these solutions are stable over astrophysical relevant timescales. Second, even if this turns to be the case, a sound formation mechanism is necessary in order to understand the possible origin of such hairy BH. Finally, if these two conditions are fulfilled, one would also need an explanation of why such kind of BH have not been observed so far. Alternatively, since there is no perfect observational data (i.e. a data with no error bars) one then wonders what would be then the largest amount of hair compatible with the current observations related with BHs, like those mentioned above. Moreover, if the forthcoming GW detections by the current and future GW instruments presents departures from the Kerr hypothesis, one can also wonder if this kind of hair can be accounted to explain these eventual anomalies.

Now, as regards the stability aspect of hairy BH, there is a long debate since the first counterexamples to the no-hair conjecture were reported in the literature \cite{hairyBH} (for a review see \cite{Volkov1999,CQGfocushairyBH}), for instance, the 
stability of {\it colored} BH, i.e., BH in the Einstein-Yang-Mills system, and other type of hairy BHs
(cf. Refs. \cite{hairyunstable,Nucamendi2003}). 
Although it had already been argued that RHBHs in the Einstein–Klein–Gordon system might be unstable
due to the presence of an ergoregion~\cite{Herdeiro2014b}, a numerical study of linear perturbations on the RHBH background, together with an analysis of the associated timescales confirming this suspicion, was not carried out until~\cite{Ganchev2018}. A subsequent analysis confirmed those conclusions~\cite{Degollado2018}. However, given that the instabilities develop on extremely long timescales $\mu t \sim 10^{11}$ for the background configurations analyzed in
\cite{Ganchev2018,Degollado2018}, the authors of Ref.\cite{Degollado2018}
termed the solutions as {\it effectively stable}, which is the terminology we will also adopt in this manuscript. For instance, they concluded that for supermassive
BHs of the order $10^9 M_\odot$, such as those observed at the centers of some galaxies, the development of these instabilities could take a timescale comparable to the age of the universe.

On the other hand, as concerns the second issue, i.e., the formation scenario, the simplest one is based on the famous {\it superradiance} instability of Kerr BH in the presence of some 
fields (the analogue of the Penrose process for extraction of energy and angular momentum when some fields are present around a rotating BH and interact gravitationally with the 
ergoregion) \cite{Wald1984,Friedman78,Brito2020a,Brito2020b}. In particular, recent full nonlinear numerical simulations show that RHBH in the Einstein-Proca system (with vector boson hair) can form by superradiance, and that the resulting hairy BHs are also stable over long timescales~\cite{East-Pretorius2017} (see also~\cite{Herdeiro-Radu2017} for further analysis and an analytical model reproducing this dynamical formation). Although one might expect that the mere presence of ergoregions could make these Proca hairy black holes unstable on sufficiently long timescales, this possibility requires further scrutiny. Nevertheless, for the scalar boson scenario, which is the one of interest in this work, such simulations have not yet been carried out, to the best of our knowledge. More precisely, the superradiant instability timescale for a scalar field around a Kerr black hole appears to be several orders of magnitude longer than in the vector case, and therefore has not yet been observed numerically.  While the origin of this striking difference between scalar and vector bosonic hair remains not fully understood, heuristic arguments based on perturbation theory suggest that the spin–spin coupling between the Kerr black hole and the vector field enhances the instability significantly. In contrast, such coupling is absent in the scalar case, resulting in much slower growth rates \cite{Brito2020a,Brito2020b}. More specifically, it appears that the spin carried by the vector boson field excites the lower superradiant modes associated with $l=0$, as opposed to the
scalar case which predominantly excites modes with $l \geq 1$. As a result, the lower centrifugal barrier of the effective potential for linear perturbations lies closer to the BH in the vector case, allowing the field to flow more rapidly across the horizon than in the scalar scenario. This, in turn, leads to faster growth of the vector field outside the BH compare to the scalar case \cite{referee}.

In view of this complicate and intricate state of affairs regarding the no-hair conjecture and the analysis of hairy BH solutions, the goal of this paper is to shed some light on the stability properties of RHBH. To investigate this issue, we perform fully nonlinear numerical simulations of different RHBH configurations. A perturbation is introduced into the stationary and axisymmetric RHBH initial data, and the subsequent evolution is followed over a sufficiently long timescale $\mu t\sim 10^3$. We then establish some criteria of (nonlinear) stability depending of the properties of the initial RHBH configuration. In particular, we observe that the configurations with more mass in the black hole than in the scalar field torus are stable within the
aforementioned time scale. The computation cost of these simulations makes it prohibitive to extend the evolutions by even an order of magnitude, let alone to reach the superradiant instability timescale $\mu t\sim 10^{11}$ reported in Refs.~\cite{Ganchev2018,Degollado2018}. Finally, we study the correlations between the oscillation modes that develop in the scalar field during the evolution following the initial perturbation.

The paper is organized as follows. Section II presents the general formalism and summarizes the construction of the initial data. Section III reviews previous results on the stability of rotating hairy black holes. Section IV is devoted to the results obtained from the numerical simulations of several equilibrium configurations with a different amount of mass in the scalar field torus. Finally, Section V concludes with a discussion of the results.

\section{Equilibrium configurations of RHBH}

\subsection{Formalism}
\label{sec:formalism}

We assume the action functional associated with GR and a matter contribution from a massive complex-valued scalar field $\Phi$:
\begin{eqnarray}
	\label{ac}
	S[g_{ab},\Phi] &= &\int\left\lbrace\frac{R}{16 \pi} - \left[\frac{1}{2}(\nabla_c \Phi^*)(\nabla^c \Phi) + \frac{1}{2}\mu^2\Phi^*\Phi\right]\right\rbrace\
	\nonumber \\
	 && \times \sqrt{-g}d^4x
\end{eqnarray}
where $\Phi^*$ stands for the complex conjugate and $\mu$ represents the scalar-field (hereafter boson field) mass. Throughout this work, we adopt geometrized units such that $G=c=1$.

The equations of motion obtained from the variation of this functional with respect to the metric and the scalar field leads
respectively to the Einstein's field equations with an energy-momentum tensor describing the distribution of the scalar-field, and the Klein-Gordon equation for the scalar field:
\begin{eqnarray}
	\label{EFE}
	&& G_{ab} = R_{ab}- \frac{1}{2} g_{ab} R= 8 \pi T_{ab} \,,\\
	\label{EMT}
	&& T_{ab} = \nabla_{(a} \Phi^* \nabla_{b)} \Phi - g_{ab}\Big[\frac{1}{2}(\nabla_c \Phi^*) (\nabla^c \Phi) +
	  \frac{1}{2}\mu^2\Phi^*\Phi \Big]\nonumber \\\\
	&& \Psi \equiv  g^{cd}\nabla_c \nabla_d \Phi - \mu^2\Phi = 0 \,.
	\label{KG}
\end{eqnarray}
Notice that we have defined the Klein-Gordon equation as $\Psi$ for later convenience.

In addition to these equations of motion, Noether’s theorem guarantees the existence of conserved quantities in this system. In particular, the invariance of the Lagrangian appearing in the action~(\ref{ac}) under global $U(1)$ transformations, $\Phi \rightarrow \Phi e^{i\alpha}$, gives rise to the conserved current
\begin{equation}
	{N_a} = {i \over 2}( \Phi^* {\nabla _a}\Phi - \Phi {\nabla _a} \Phi^*).
\end{equation} 
that satisfies $\nabla_a N^a = 0$. The spatial integral of the time component of this current defines a conserved quantity, known as the Noether charge $N$, which is of particular interest because it can be directly associated with the total amount of bosonic matter present in the system.

\subsection{General assumptions of RHBH}
\label{sec:general}

We plan to construct initial data representing rotating hairy black holes (RHBH), which are stationary, axisymmetric, asymptotically flat and without ``meridional currents", i.e., under the circularity condition. We thus assume the existence of two killing vector fields, $\xi^a$ and $\eta^a$, which are, respectively, time-like and space-like, in the asymptotic regions, and are associated with the hypothesis of stationarity and axisymmetry. These Killing fields commute and
are taken as part of the coordinate basis. We use the time $t$ and the spatial angle $\varphi$ as coordinates adapted to these vector fields
\be
\label{KF}
\xi^a  = \left(\frac{\partial}{\partial t}\right)^a \,\,\,,\,\,\,\eta^a = \left(\frac{\partial}{\partial\varphi}\right)^a \,.
\ee

In the seminal works on RHBHs~\cite{Herdeiro2014, Herdeiro2015}, the Einstein–Klein–Gordon system was solved numerically using a finite-difference method.
These solutions were later reproduced using spectral methods \cite{Garcia2023} implemented in the KADATH library \cite{Grandclement2010}, adopting a gauge based on spherical-like, quasi-isotropic coordinates (QIC).  Those studies also reported a large set of global quantities \cite{Garcia2023}. In the investigations of \cite{Herdeiro2014, Herdeiro2015, Grandclement2009}, the use of spherical-like coordinates adapted to the spacetime symmetries, together with the circularity condition, greatly simplifies the gravitational sector: several metric components vanish identically throughout the domain outside the black hole, including on the horizon.
The ansatz for the scalar field is also relatively straightforward. However, some metric quantities (i.e., like the lapse function) vanish at the black hole horizon, since in this case the latter coincides
with the Killing horizon associated with the helical Killing field
$\chi^a= \xi^a + \Omega_{\rm BH} \eta^a$, where $\Omega_{\rm BH}$ is the angular velocity of
the BH. Furthermore, the spatial coordinates
adapted in ~\cite{Herdeiro2014, Herdeiro2015} are singular at the BH horizon, where the component
$g_{rr}$ diverges there (i.e., like in the Boyer-Lindquist coordinates for the Kerr solution), reflecting the fact that
these coordinates are not regular across that surface. However, this is not the case of
the QIC employed in \cite{Garcia2023}, although these coordinates do not allow to penetrate the horizon
either. Such lack of regularity can pose difficulties, particularly when attempting to evolve such configurations dynamically. 

A different strategy was adopted in~\cite{Grandclement2022}, where a formalism based on maximal slicing and the spatial harmonic gauge was introduced and applied to various spacetimes containing black holes, including the RHBHs studied in this paper. This gauge choice, a generalization of the Eddington–Finkelstein coordinates, has the advantage of remaining regular across the horizon. Its application to the computation of RHBH solutions is described extensively in Sec.~V of~\cite{Grandclement2022}. Here, we only summarize the main aspects and refer the interested reader to that work for further details. Note, however, that we employ a different notation from the one used there.

The spacetime metric is decomposed using the standard 3+1 formalism into the lapse function $\alpha$, the shift vector $\beta^i$ and the spatial metric $\gamma_{ij}$. All these quantities, given the symmetries of the system, only depend on the coordinates $\l(r, \theta\r)$. The full spacetime line element relates to the 3+1 quantities by the usual form: 
\be
\label{3p1}
{\rm d}s^2 = -\l(\alpha^2 - \beta^i \beta_i\r) {\rm d}t^2 + 2 \beta_i {\rm d}x^i {\rm d}t + \gamma_{ij} {\rm d}x^i {\rm d}x^i.
\ee

In the formalism developed in~\cite{Grandclement2022}, the numerical unknowns are the 3+1 fields themselves, rather than selected metric components as in the quasi-isotropic  case. These fields are obtained by solving the full set of 3+1 Einstein equations: the Hamiltonian constraint, the momentum constraints, and the evolution equation for the spatial metric. The presence of a horizon is enforced by solving the equations only outside a spherical surface, which is defined as being the horizon. On this surface, a set of nontrivial boundary conditions is imposed, derived from the requirement that the surface represents an apparent horizon in equilibrium. Certain quantities on the horizon can be freely specified as coordinate choices (like for instance, the value of the lapse function). Finally, the degeneracy of some equations on the horizon provides the remaining information needed to ensure a well-posed problem.

In this context, the complex scalar field is expressed as
\be
\label{ansatz}
\Phi \l(t,r,\theta,\varphi \r) = \phi (r, \theta) \exp \l[i\l(\omega t - m \varphi\r)\r],
\ee
where $\phi = \phi_R + i \phi_I$ depends only on $\l(r, \theta\r)$, with $\phi_R$ and $\phi_I$ being real-valued functions. The positive integer $m$ denotes the winding number, while $\omega$ is the angular frequency of the scalar field. Unlike the QIC case
\cite{Garcia2023}, the imaginary component $\phi_I$ does not vanish. The Klein–Gordon equation~(\ref{KG}) preserves the same harmonic dependence as the ansatz in Eq.~(\ref{ansatz}), and can thus be written as
\be
\label{KG_ansatz}
\Psi\l(t,r,\theta,\varphi \r) =  {\rm \psi} (r, \theta) \exp \l[i\l(\omega t - m \varphi\r)\r].
\ee
with $\psi(r, \theta) = {\rm \psi}_R + i{\rm \psi}_I$. The two real equations ${\rm \psi}_R(r, \theta) = 0$ and ${\rm \psi}_I(r, \theta) = 0$ must both be satisfied to obtain a complete solution for the scalar field. It is worth noting that these equations become degenerate on the horizon, implying that the scalar field’s value there cannot be freely prescribed but is instead determined self-consistently by the system. In particular, the solution depends on the parameters $\mu$, $m$, and $\omega$.

A subtlety arises from the fact that there is a constant phase freedom in the definitions of $\phi_R$ and $\phi_I$. A valid procedure for fixing this phase is discussed in \cite{Grandclement2022}. The important implication is that solutions of the full system can only exist if the angular velocity of the horizon, $\Omega_{\rm BH}$, and the frequency of the field, $\omega$, satisfy the so-called synchronization condition
\be
\label{synchronization}
\omega = m \Omega_{\rm BH}.
\ee
which requires that, for a stationary solution, the scalar field oscillates as an multiple integer
of the BH's angular velocity. Here we focus only on initial data representing RHBH with $m=1$.

\subsection{Global quantities}
\label{sec:global}

The equilibrium configurations can be characterized by their mass and angular momentum, both those of the black hole and of the surrounding scalar field torus. As in \cite{Grandclement2022}, the total mass and angular momentum of each configuration are computed through surface integrals at infinity. More precisely, the ADM mass is given by
\be
\label{e:def_adm}
M = \frac{1}{16 \pi}\int_{r=\infty} f^{ik}  f^{jl}  \l(\hat{D}_j \gamma_{kl} - \hat{D}_k \gamma_{jl}\r) {\rm d}S,
\ee
where $\hat{D}$ denotes the covariant derivative associated with the flat metric $f_{ij}$ and ${\rm d}S$ is the surface element at infinity.

The angular momentum with respect to the $z$-axis is obtained from
\be
\label{e:def_momentum}
J =  \frac{1}{8 \pi} \int_{r=\infty} K_{ij} m^i s^j {\rm d}S.
\ee
where $K_{ij}$ is the extrinsic curvature tensor, $s^i$
the normal radial vector and $m^i$ denotes the rotation vector around the $z$-axis.

Since the spacetime is stationary, these quantities must coincide with their Komar counterparts, which are computed from the stress-energy tensor of the scalar field given by Eq.~\eqref{EMT}. 
The computation of the Komar mass consists of two contributions. 
The first one, $M_{\rm BH}$, involves a surface integral on the horizon and reflects the presence of a black hole. The second term, $M_\Phi$, contains a volume integral on some components of the stress-energy tensor and is associated with the presence of the scalar field. Using the decomposition described above, one then finds
\bea
\label{e:mbh}
M_{\rm BH} &=& \frac{1}{4\pi} \int_{r=r_{\rm h}} \sqrt{h} \l({s}^i D_i \alpha - K_{ij} {s}^i \beta^j\r) {\rm d} S \\
\label{e:mphi}
M_\Phi &=& \int_{\Sigma_t} \sqrt{\gamma} \l[\alpha \l(E+\gamma^{ij} S_{ij}\r) - 2 S_i \beta^i\r] {\rm d}^3 x,
\eea
where $h$ denotes the determinant of the metric induced on the horizon $h_{ij} = \gamma_{ij} - {s}_i {s}_j$ and $D$ denotes the standard covariant derivative associated with $\gamma_{ij}$. Here we have
decomposed the stress-energy tensor as measured by an  observer moving along the vector $n_a = (-\alpha, 0,0,0)$, namely
\begin{equation}
	T_{ab} = E n_a n_b + S_a n_b + S_b n_a + S_{ab}
\end{equation}
where $E$, $S_a$, and $S_{ab}$ represent, respectively, the energy density, momentum density, and stress (pressure) tensor of the scalar field as measured by this observer. These quantities satisfy the orthogonality conditions $S^a n_a = 0$ and $S^{ab} n_b = 0$.

The situation is similar for the angular momentum, which can also be decomposed into two contributions: $J_{\rm BH}$, a surface integral over the horizon providing a local measure of the black hole’s spin, and $J_\Phi$, a volume integral involving the matter terms. It is worth noting that $J_\Phi$ is related to the Noether charge $N$ of the scalar field via the simple relation $J_\Phi = m N$. Following \cite{Grandclement2022}, within the context of this work, these quantities can be expressed as
\bea
\label{e:noether}
J_{\rm BH} &=& \frac{1}{8\pi} \int_{r=r_{\rm h}} \sqrt{h}\,  K_{ij} m^i {s}^j {\rm d} S, \\
J_\Phi &=& -m \int_{\Sigma_t} \frac{1}{\alpha} \l[\l(\omega + m \delta_i^\varphi \beta^i\r) \l(\phi_R^2 + \phi_I^2 \r)\r. \\
\nonumber
&+& \l. {\beta^i}\l(\phi_I D_i {\phi}_R - \phi_R D_i {\phi}_I \r)\r] \sqrt{\gamma} {\rm d}^3 x.
\eea

The accuracy of the configurations can be assessed by computing the relative differences between the ADM mass and total angular momentum and their Komar counterparts. 
Using the standard resolution employed here (17 coefficients in both the $r$ and $\theta$ directions), the relative differences in the masses are of order $10^{-4}$. A similar level of accuracy is observed for the angular momentum.

\subsection{Numerical construction of the initial data}
\label{sec:numerical_ID}

The Einstein-Klein-Gordon system, under the symmetries described in Section \ref{sec:general}, is solved using the KADATH library \cite{Grandclement2010} that enables the use of spectral methods to solve partial differential equations arising in general relativity and theoretical physics \cite{Grandclement2009}. The unknown fields are the 3+1 quantities and the two components $\phi_R$ and $\phi_I$ of the scalar field. These fields are governed by the Einstein-Klein-Gordon system, which is solved only outside the horizon. The physical space is divided into several (typically ten) spherical domains. The system is closed by imposing apparent horizon boundary conditions at the inner boundary and asymptotic flatness at spatial infinity, the latter enforced exactly through a compactification of spacetime using the variable $1/r$.

The nonlinear system of equations is solved iteratively using the Newton-Raphson method. As a first step, the so-called cloud solution is obtained, in which the metric is fixed to that of a Kerr black hole and only the Klein-Gordon equation for the scalar field is solved. In this particular work, the coordinate horizon radius is $1/\mu_{\rm{fid}}$, the black hole mass is $M_{\rm BH} \approx 1.73/\mu_{\rm{fid}}$ and the Kerr parameter is $J_{\rm BH}/M^2_{\rm BH} \approx 0.5$. Here $\mu_{\rm{fid}}$ is a {\it fiducial} mass unit that is introduced for numerical purposes, but at the end of the calculations we rescale conveniently all quantities and report them in units of $\mu$. The value of the BH mass results from the choice of freely specifiable quantities on the horizon, in particular the lapse (see \cite{Grandclement2022}). With this setup, the angular velocity of the black hole is $\Omega_{\rm BH} = 0.08 \mu_{\rm{fid}}$. Using this Kerr spacetime as a background, the cloud solution can then be computed, yielding a scalar field mass of $\mu \approx 0.0802\, \mu_{\rm{\rm{fid}}}$.

Such cloud solution can serve as an initial guess for the solver to obtain configurations of the full coupled Einstein-Klein-Gordon system. To achieve this, one must move away from the parameter values of the cloud solution. In \cite{Grandclement2022}, this is done by computing solutions with different values of $\Omega_{\rm BH}$ (see for instance Fig. (13) of \cite{Grandclement2022}). In the present work, however, $\Omega_{\rm BH}$ is kept fixed, and a sequence of solutions is generated by varying slightly the value of $\mu$. This procedure allows us to obtain configurations in which the scalar field mass $M_\Phi$ increases relative to that of the black hole $M_{BH}$.

\begin{table}[h]
	\centering
	\begin{tabular}{|c|c|c|c|c|c|c|}
		\hline
		& $\mu M$ & $\mu^2 J$ & $M_{\Phi}/M$ & $M_{\rm BH}/M$ & $J_{\Phi}/J$ & $J_{\rm BH}/J$ \\ \hline
		RHBH04   & 0.143   & -0.015   & 0.04 &  0.96 & 0.34 & 0.66 \\ \hline
		RHBH20   & 0.174   & -0.045   & 0.20 & 0.80 & 0.79 & 0.21 \\ \hline
		RHBH28  & 0.192  & -0.063  & 0.28 & 0.72  & 0.85 & 0.15 \\ \hline
		RHBH34  & 0.210  & -0.081  &  0.34 & 0.66 & 0.88 & 0.12 \\ \hline
		RHBH50  & 0.276  & -0.145  & 0.50  &  0.50 & 0.93 & 0.07 \\ \hline
		RHBH68  & 0.452  & -0.315 &  0.68 & 0.32 & 0.96 & 0.04 \\ \hline		
	\end{tabular}
	\caption{Summary of the RHBH configurations, denoted with the fraction of mass in the scalar field with respect to the total one. The mass of the black hole remains fixed to $M_{\rm BH} \approx 1.73/\mu_{\rm{fid}}$ and its Kerr parameter is $J_{\rm BH}/M^2_{\rm BH} \approx 0.5$. The mass of the scalar field particle for these configurations is $\mu \approx 0.0802\, \mu_{\rm{\rm{fid}}}$, such that $\mu M_{\rm BH} \approx 0.138$.
  } 
	\label{tab:RHBHconfiguration}
\end{table}

\begin{figure}
	\centering
	\includegraphics[width=0.9\linewidth]{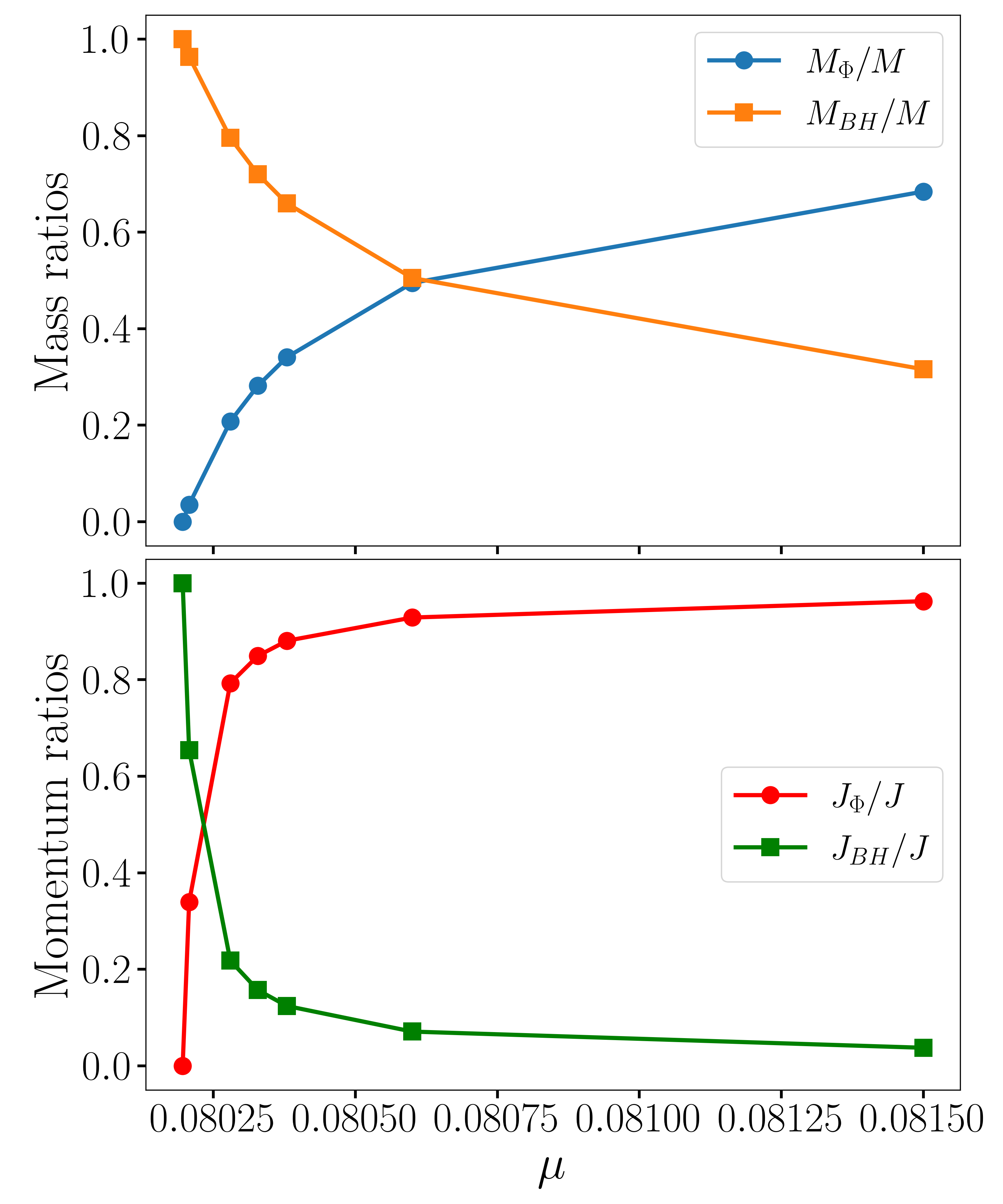} 
	\caption{\label{fig:global} {\em Global properties of the RHBH configurations.} Relative contribution of the horizon and of the scalar field  to the mass (top) and to the angular momentum (bottom). The leftmost solution represents Kerr background with no scalar field.}
\end{figure}

The relative contributions of the scalar field and the black hole to the total mass can be quantified by plotting $M_\Phi / M$ and $M_{\rm BH} / M$, whose sum is always equal to one. A similar decomposition can be applied to the angular momentum by considering $J_\Phi / J$ and $J_{\rm BH} / J$. These ratios are shown in Fig.~\ref{fig:global} and reported explicitly in Table~\ref{tab:RHBHconfiguration}. As mentioned previously, the sequence originates from a cloud solution with $\mu \approx 0.0802\, \mu_{\rm{\rm{fid}}}$ on a Kerr background with $J_{\rm BH}/M^2_{\rm BH} \approx 0.5$. Configurations with a larger scalar-field contribution are obtained by increasing $\mu$ while keeping $\Omega_{\rm BH}$ fixed.

\subsection{Interface between the initial data and evolution code}
\label{sec:interface}

The procedure for transferring data from KADATH to the evolution code proceeds as follows. The initial data are computed numerically using a spherical orthonormal tensor basis, whereas the evolution code requires all tensor quantities to be expressed in a Cartesian basis. As a first step, the tensor basis is transformed using the appropriate conversion formulas, which primarily involve multiplications by trigonometric functions. These operations are efficiently performed within the KADATH framework \cite{Grandclement2010}, which employs a spectral representation of the fields.

The evolution code handles the divergences present in the black hole interior using the so-called puncture method~\cite{2007PhRvL..99x1102H}. This approach requires the field values to be initially specified over the entire computational domain, including the region inside the horizon. This contrasts with the initial data solver, which considers the EKG system only outside the apparent horizon. To address this discrepancy, all fields are smoothly extended inside the horizon using second-order Lagrange interpolants. This technique has already demonstrated its effectiveness, for instance, in the context of the FUKA initial data solver \cite{Papenfort2021}.

The procedure is quite straightforward and proceeds as follows. The evolution code first defines a grid structure covering the entire computational domain, possibly including multiple refinement levels. The grid points of the evolution code are then provided to the spectral solver, which evaluates the field values at each point through spectral summation, ensuring spectral accuracy. The resulting values of all relevant fields are subsequently transferred to the evolution code. Specifically, the required gravitational quantities are the lapse $\alpha$, the components of the shift vector $\beta^i$,
the spatial metric tensor $\gamma_{ij}$ and the extrinsic curvature tensor $K_{ij}$. As before, all tensors are expressed in a Cartesian tensor basis. The real and imaginary parts of the scalar field $\Phi$ must also be given. They are constructed from $\Phi_R$ and $\Phi_I$ using Eq.~(\ref{ansatz}) evaluated at $t=0$. Finally, the real and imaginary parts of the time derivative of the scalar field, also derived from Eq.~(\ref{ansatz}), provide the remaining necessary information.

\section{Previous stability studies of RHBH}

It is important to provide a brief review of the known stability properties of these and other related hairy black holes (e.g., rotating Proca black holes), both to highlight their differences and to relate them to the nonlinear stability properties investigated in this work.

As mentioned earlier in the Introduction, after the discovery of RHBH by Herdeiro \& Radu \cite{Herdeiro2014,Herdeiro2015}, it was argued that 
some of these solutions (with $m=1$ and scalar-field perturbations having ${\tilde m}=2$) are unstable~\cite{Ganchev2018}. Subsequent work, however, put that conclusion in a more general context~\cite{Degollado2018}. For instance, a careful comparison of the RHBH existence regions (see Fig. 1 of Ref.~\cite{Degollado2018} and Fig. 12 of Ref.~\cite{Garcia2023}) with the parameter region explored in \cite{Ganchev2018} reveals that the latter covers only a minuscule portion of the full parameter space. Consequently, it is premature to draw definitive conclusions about the stability of the much larger region left unexplored in \cite{Ganchev2018}.
The small region studied in \cite{Ganchev2018} lies close to the existence line of cloud solutions around a Kerr BH for 
$m=1$; moreover, the background unperturbed RHBH solutions
considered there range from nearly extremal ($J/M^2\sim 0.95$) to superextremal ($J/M^2>1$) regions.
Since these configurations lie close to the cloud solutions, most of the mass and angular momentum is provided by the black hole. By contrast, the configurations we study here have moderate BH angular momentum $J_{\rm BH}/M^2_{\rm BH} \approx 0.5$, and therefore probe a different and much larger portion of the solution space.

Moreover, as shown in Ref.~\cite{Ganchev2018}, the imaginary part $\tilde \omega_I$ of the complex-valued frequency $\tilde \omega$, associated with the scalar-field perturbation\footnote{Notice that 
the frequency of the background scalar hair is denoted by $\omega$, while $\tilde \omega$ is the frequency of the scalar-field perturbation. This is the opposite convention used in Ref.\cite{Ganchev2018}.}, is of the same order 
as $\tilde \omega^{\rm Kerr}_I$, which corresponds to the imaginary part of the perturbation frequency around Kerr. This imaginary component is particularly relevant, as it determines the characteristic timescale of the instability. Since Kerr black holes are susceptible to superradiant instabilities, driven by the ergoregion and affecting matter perturbations such as scalar fields, this mechanism is a natural route for the formation of at least some RHBHs. In the Kerr case, for the region analyzed in  Ref.~\cite{Ganchev2018}, one finds $\tilde \omega^{\rm Kerr}_I \sim 10^{-11} \mu$,  which corresponds to an exceedingly long characteristic timescale $t\mu \sim {\cal O}(10^{11})$ for RHBH  with total ADM mas $M\mu \lesssim 1$.
In view of these extremely long timescales, the instability region identified in \cite{Ganchev2018} was termed {\it effective stabililty} in \cite{Degollado2018}. This instability is presumably still related to superradiance, triggered by the presence of an ergoregion in the RHBH. Its characteristic timescale was found to be of the same order of magnitude as the superradiant instability of a Kerr BH, according to the results reported in Refs.~\cite{Ganchev2018,Degollado2018}.
This explains why it has so far been impossible to confirm, by using fully non-linear evolutions of the EKG system, whether some RHBHs are indeed the end states of Kerr superradiant instabilities. The situation contrasts with the vector-boson (Proca) case, for which this formation mechanism was confirmed by nonlinear simulations \cite{East-Pretorius2017}. In the Proca scenario the instability grows three to four orders of magnitude faster than in the scalar case. As it was mentioned before, the spin-1 nature of the vector field couples to the black-hole rotation in a way that strongly enhances the instability. Once a hairy Proca black hole forms (i.e. when the synchronization condition is reached), the system relaxes to a nearly stationary configuration, and no further instability is observed over the long evolutions reported in \cite{East-Pretorius2017} (see also \cite{East2017}).
Note that, as emphasized in the Introduction, these studies do not exclude the presence of long-term superradiant instabilities.
Although scalar RHBH solutions have not yet been produced dynamically through fully nonlinear evolutions triggered by superradiant instabilities, they can be constructed non-dynamically using the procedure described in the previous section, and their stability can be investigated through nonlinear time evolutions. This has been achieved very recently in an independent and simultaneous study by Nicoules {\it et al.}~\cite{2025arXiv250920450N}, who employed a completely different numerical framework for both the construction of the initial data and the subsequent evolutions compared to the methods used in this work.
For the construction of the initial data, they employed finite-difference methods, while the time evolutions were performed using the {\tt Einstein Toolkit} infrastructure within the BSSN formalism. They considered three representative initial configurations, covering a range from low to high scalar-hair content: $M_{\rm BH}/M \simeq 0.75$, $0.12$, and $0.04$, corresponding to scalar-field mass fractions of approximately $25\%$, $88\%$, and $96\%$ of the total ADM mass $M$, respectively. For the low hair-mass configuration, they found that the system remains stable over the simulated timescale, whereas the two other  configurations exhibit clear signs of instability.

As we show in the next section, our results are consistent with the conclusions of Ref.~\cite{2025arXiv250920450N}.
In our study, we consider six initial RHBH configurations with different amounts of scalar hair. We typically observe that configurations with $M_\Phi/M > 0.5$ and moderate BH angular momentum become unstable. We perform a mode decomposition in the azimuthal angle of the scalar field torus and find that the instability of the RHBH is likely associated with the non-axisymmetric instability (NAI) recently identified in rotating boson stars~\cite{NAI1,NAI2,NAI3}. Another consequence of this instability is that the black hole drifts outward following a spiral-like trajectory in the $x$–$y$ plane, moving along the direction of the system’s rotation. Upon reaching the scalar-field torus, the black hole begins to disrupt its structure by accreting part of the surrounding scalar matter.

The remaining configurations with $M_\Phi/M \lesssim 0.5$ remain stable over the timescales explored in our simulations. In these cases, the central black hole exhibits only small oscillations around its equilibrium position, while the scalar field configuration remains stable and axisymmetric throughout the entire evolution.

\section{Results of the numerical simulations}
\label{sec:numanalysis}

In this section, we present the results of our simulations of RHBH configurations, conducted to assess their stability and explore their dynamical behavior. After a brief overview of the numerical methods and grid setup, we present the results from simulations of six different RHBH configurations, explicitly listed in Table~\ref{tab:RHBHconfiguration}. We then provide a qualitative description of the dynamics, followed by a quantitative analysis. Finally, we discuss the stability properties of the configurations, demonstrating that some RHBH solutions are indeed unstable.

\subsection{Numerical methods, evolution equations and grid setup}
\label{sec:setup}

The numerical simulation was performed using the code {\sc MHDuet}~\cite{mhduet_webpage}. {\sc MHDuet} is automatically generated by the platform {\sc Simflowny} \cite{arbona13,arbona18} to run on the {\sc AMReX} infrastructure, which provides parallelization both in CPU/GPUs and adaptive mesh refinement. It uses fourth-order-accurate operators for the spatial derivatives in the EKG equations, supplemented with sixth-order Kreiss-Oliger dissipation; a fourth-order Runge-Kutta scheme with sufficiently small time step $\Delta t \leq 0.2~\Delta x$ (where $\Delta x$ is the grid spacing); and an efficient and accurate treatment of the refinement boundaries when sub-cycling in  time~\cite{McCorquodale:2011,Mongwane:2015}. 
A complete assessment of the accuracy and robustness of the numerical methods implemented can be found in \cite{palenzuela18,palenzuela25}.

We adopt the covariant conformal Z4 (CCZ4) formulation~\cite{2017PhRvD..95l4005B,Alic12}. The Z4 formulation~\cite{Bona03} extends Einstein’s equations by introducing a four-vector that measures deviations from exact Einstein solutions. These covariant equations are then decomposed using the 3+1 formalism and evolved in terms of conformal quantities. The system is supplemented with standard gauge conditions, namely the 1+log slicing condition for the lapse and the Gamma-driver condition for the shift. A detailed description of the evolution equations and gauge choices can be found in Ref.~\cite{2017PhRvD..95l4005B}. We want to stress that this code, with basically the same equations and numerical schemes, has been employed successfully to study the dynamics of highly compact boson stars~\cite{2017PhRvD..95l4005B, 2017PhRvD..96j4058P,2022PhRvD.105f4067B}. 

The RHBH solutions studied here present a wide range of spatial and time scales. On one hand, the BH mass remains approximately fixed for all the configurations, setting the size of the BH horizon $R_{\rm BH} \approx M_{\rm BH}$. On the other hand, the size of the scalar field torus extends up to few hundred times the BH horizon, $R_{\phi} \approx {\cal O}(100) R_{\rm BH}$. Notice however that beyond the bulk of the torus, the scalar field still decays with a slow exponential tail that extends up to ${\cal O}(1000) R_{\rm BH}$. 

Resolving both the compact black hole and the extended scalar field torus with sufficient accuracy is challenging due to their disparate scales, although such difficulty can be efficiently addressed through fixed mesh refinement. Recall that the BH size is characterized by $\mu M_{\rm BH} \approx 0.138$. Accordingly, the RHBH configurations are evolved within a cubic computational domain of size $[-\mu L_0, \mu L_0]^3=[-192, 192]^3$ employing 10 additional refinement levels, each defined on a cubic subdomain of size $\mu L_l$. The first three refinement levels are used to fully encompass the scalar field torus, with sizes $\mu (L_1,L_2,L_3)=(112,88,64)$. The remaining levels focus on resolving the inner regions of the torus and the BH itself, with sizes $\mu (L_4,L_5,L_6,L_7,L_8,L_9,L_{10})=(5.6,1.92,0.96,0.64,0.48,0.32,0.24)$. The effective resolution achieved in the scalar field torus is $\mu \Delta x = 0.376$, corresponding to approximately 200 grid points across its diameter. The finest refinement level covering the black hole attains a resolution of $\mu \Delta x = 0.0029$, ensuring that the horizon is resolved by roughly 60 grid points.

\subsection{Dynamics of the RHBH}
\label{sec:dynamics}

\begin{figure}
	\centering
	\includegraphics[width=0.87\linewidth]{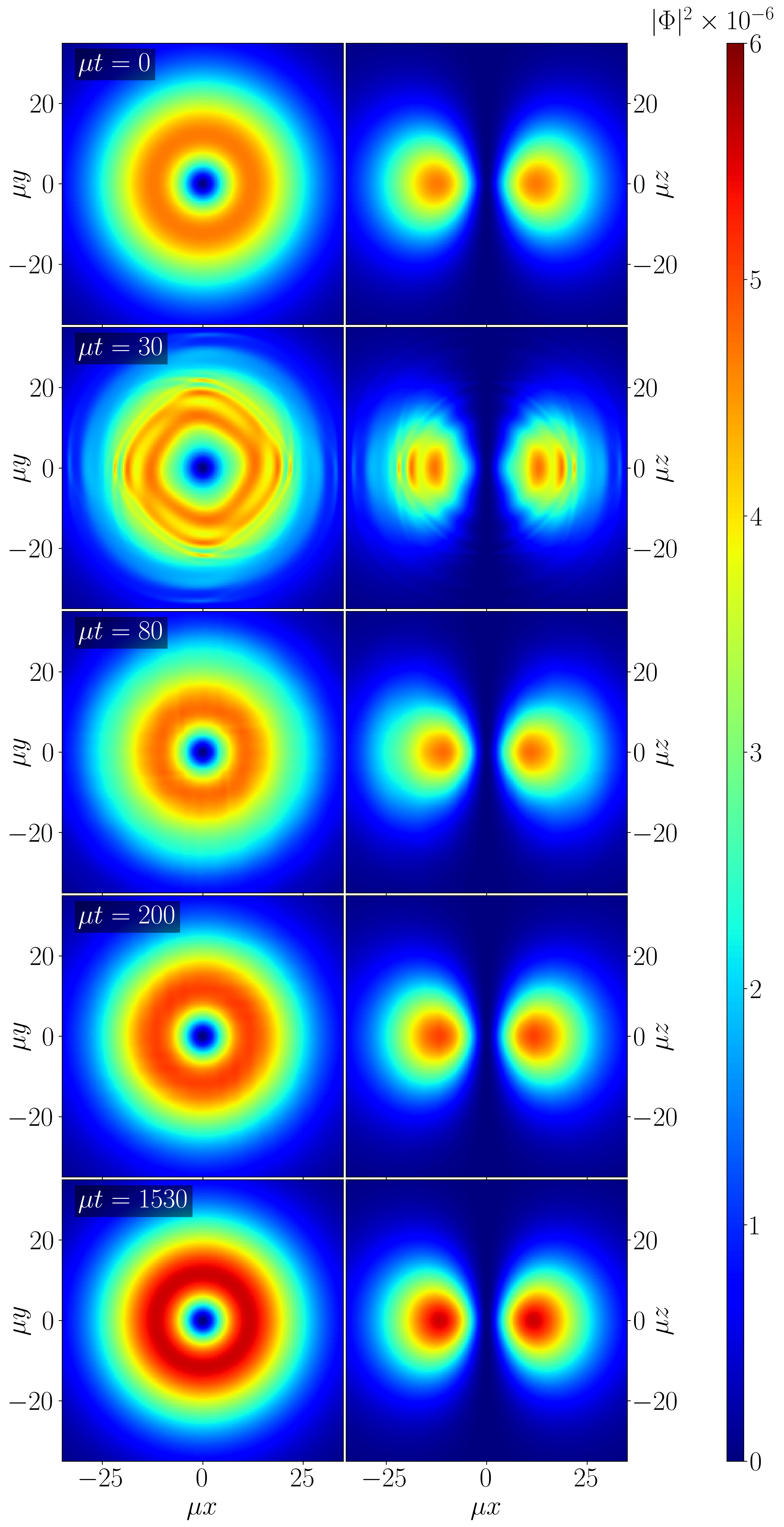}
	\caption{\label{fig:RHBH_66} {\em Dynamics of the configuration RHBH34}. Snapshots of the scalar field density $|\Phi|^2$, along the the equatorial and meridional planes for some illustrative times. At early times there is transient in the torus induced by the initial strong perturbation, which propagates back and forth until that it relaxes to  a final solution slightly different from the initial one. This new solution remains axisymmetric and stable for long timescales, at least up to $\mu t \approx 1600$.}
\end{figure}

Our initial data correspond to the RHBH solutions presented in Section~\ref{sec:numerical_ID}, with an additional perturbation introduced to trigger a non-trivial dynamical evolution and to probe the stability of these configurations on a commensurate timescale. The perturbation is implemented by intentionally enhancing numerical discretization errors at the boundaries between refinement levels (i.e., by using low-order spatial discretization operators which are second-order accurate on the faces and zeroth order at corners and edges). While most of such errors are expected to vanish in the limit of infinite resolution, in the simulations considered here the resulting perturbation has a relatively large amplitude and manifests as several cubic shells. These features break axial symmetry and may excite a broad range of modes. Although the computational cost of the simulations prevented a systematic exploration, similar qualitative behaviour is observed for alternative perturbations and across different numerical resolutions (see Appendix~\ref{appendixA} for further details).

\begin{figure}
	\centering
	\includegraphics[width=0.87\linewidth]{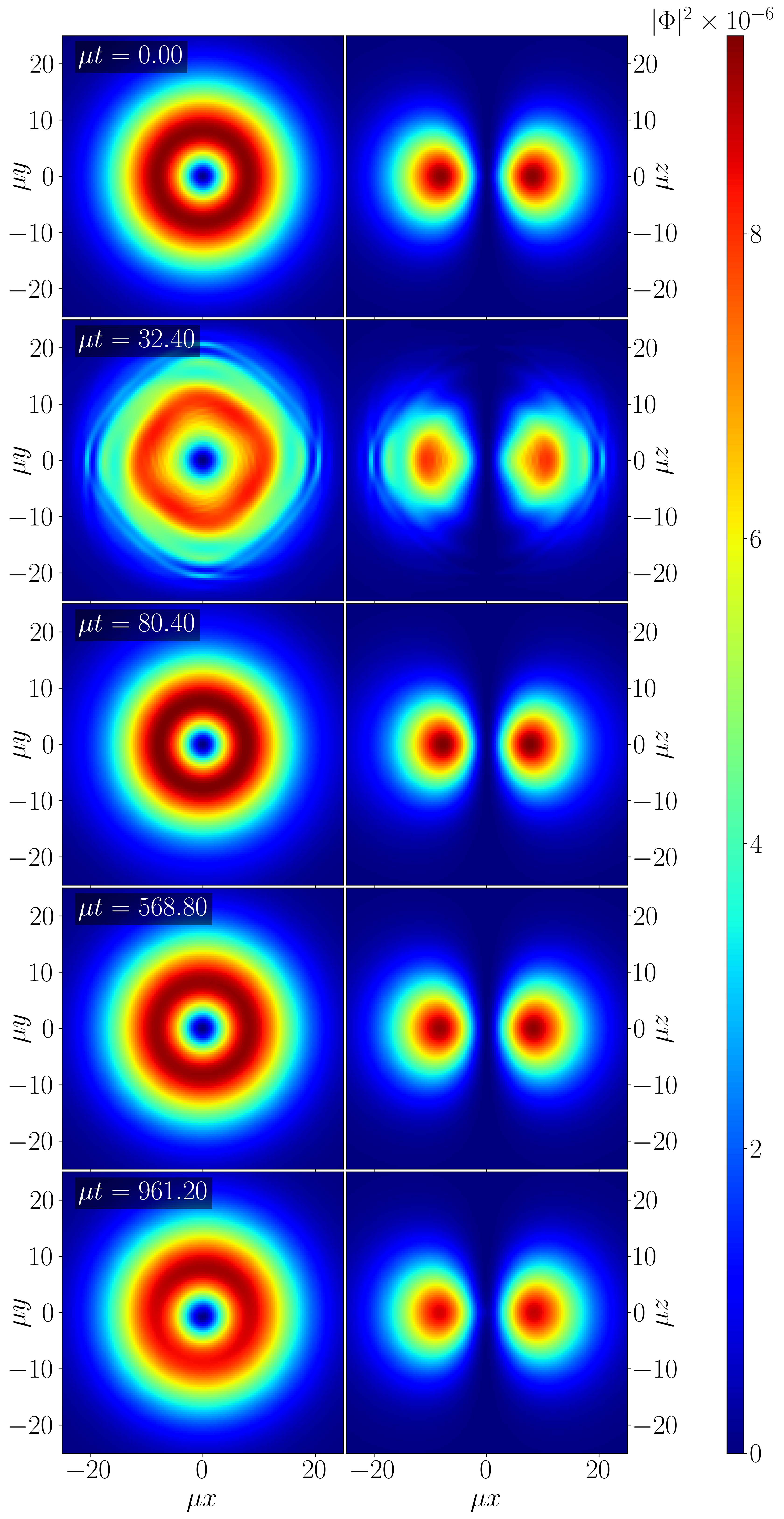} 
	\caption{\label{fig:RHBH_32} {\em Dynamics of the configuration RHBH68}. Snapshots of the quantity $|\Phi|^2$ along the equatorial and meridional planes at selected illustrative times. There is an initial transient in the torus induced by the initial perturbation, which excites non axisymmetric modes. More specifically, the instability manifests as an exponential growth for azimuthal modes $m>0$ (see section~\ref{sec:stability_analysis}) within the scalar field torus, visible only at late times (bottom row). The black hole also experiences a slight displacement from the origin, though this is not discernible at the plot’s scale.}
\end{figure}

The dynamics observed for the five RHBH configurations with the lowest scalar field masses (i.e., from RHBH04 to RHBH50) are qualitatively very similar. Therefore, we illustrate the typical behavior by presenting the particular case  RHBH34. Time snapshots of the scalar field density $|\Phi|^2$ are shown in Fig.~\ref{fig:RHBH_66}, both in the meridional and equatorial planes. A similar evolution is also observed in other quantities, such as the Noether charge density. After a violent initial transient caused by the strong perturbation, the RHBH settles into a stable configuration that differs slightly from the initial state and remains in this quasi-stationary phase for at least $\mu t \sim 1600$ (corresponding to $t/M_{BH} \sim 11000$), showing no appreciable signs of instability.

\begin{figure} 
	\centering
	\includegraphics[width=0.48\textwidth]{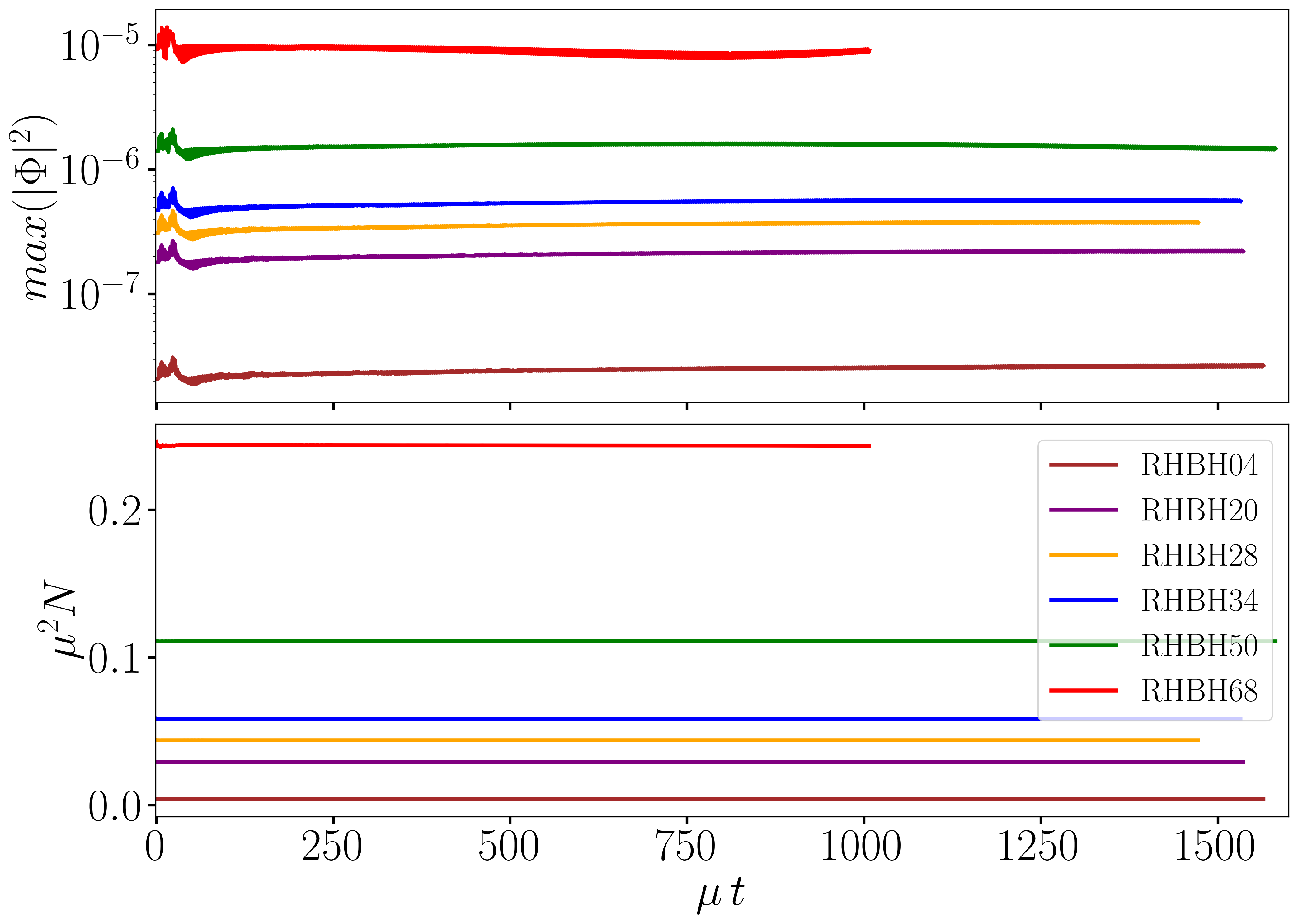} 
	\caption{{\em Dynamics of the RHBH configurations}. (Top) Maximum of $|\Phi|^2$ for all the configurations studied here. All cases exhibit an initial transient phase induced by the strong perturbation in the initial data, followed by a slight increase in 
	$\max \left(|\Phi|^2\right)$. For configurations with $M_{\Phi}/M \lesssim 0.5$, the system subsequently relaxes into a stationary, oscillatory state.  For larger mass fractions $\max \left(|\Phi|^2\right)$ grows more significantly, and the initial transient becomes more violent. A subsequent decay followed by a rebound is then observed, signaling a strong perturbation of the scalar field torus by the black hole.
	(Bottom) In all cases, the global Noether charge remains constant with high-accuracy, confirming the reliability of the simulations and indicating that the scalar field is neither escaping to infinity nor being accreted by the black hole. } 
	\label{fig:phi2max_Noether} 
\end{figure}

\begin{figure} 
	\centering
	\includegraphics[width=0.48\textwidth]{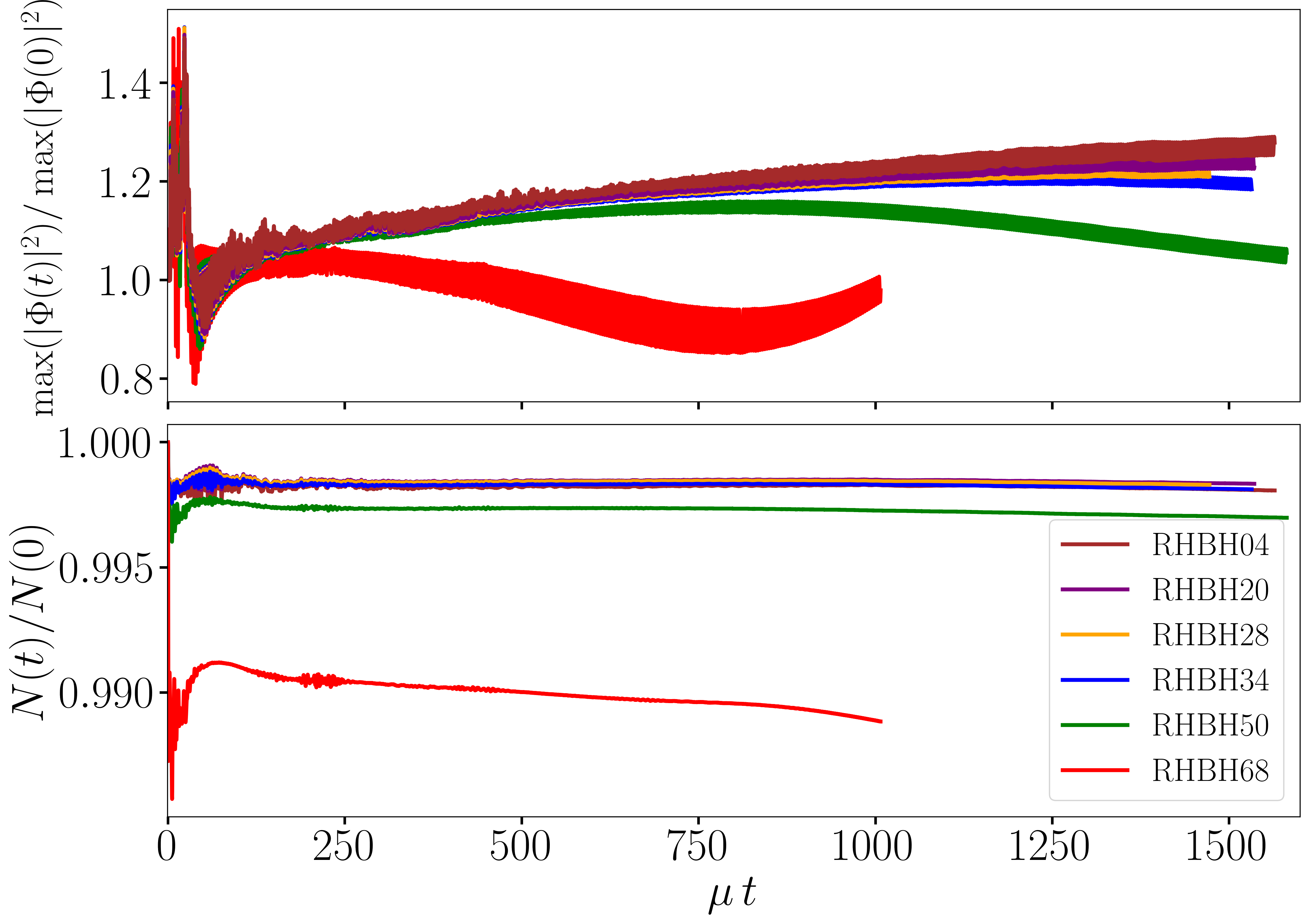} 
	\caption{{\em Dynamics of the RHBH configurations}. Same as Fig.~\ref{fig:phi2max_Noether}, but with all quantities normalized to their initial values at $t=0$. This normalization makes it easier to discern the evolution of the maximum and the good conservation of the global Noether charge.}
        \label{fig:phi2max_Noether_norm}
\end{figure}

The time evolution of the RHBH configuration with the highest scalar field mass (i.e., RHBH68) is markedly different. Again, time snapshots of $|\Phi|^2$ are shown in Fig.~\ref{fig:RHBH_32}, both in the meridional and equatorial planes. In this case, after the initial transient, the RHBH remains stable only for a short time before becoming unstable. The instability manifests in two distinct ways: first, through the growth of non-axisymmetric modes in the scalar field torus, and second, through a displacement of the black hole from the origin, following an outward spiral trajectory until it eventually collides with the torus. The presence of this instability is clearly visible at the final time shown. In particular, in the equatorial plane, a density enhancement develops in a localized region of the ring, indicating that the matter distribution undergoes a spatial rearrangement and the system departs from equilibrium, a behavior absent in the stable configurations.

A more quantitative analysis of the dynamics can be performed by examining several relevant global quantities. For the RHBH, the evolution can be characterized by the maximum of $|\Phi|^2$, the oscillation frequency of the scalar field $\omega$, and the global Noether charge $N$. As observed in the time snapshots, the fraction of the total mass contained in the scalar field has a strong influence on the final state of the RHBH, although it plays a less significant role in the evolution of the maximum scalar field density $|\Phi|^2$, shown in the top panel of Figure~\ref{fig:phi2max_Noether} (see
  also Figure~\ref{fig:phi2max_Noether_norm}). For the stable configurations, the maximum of 
$|\Phi|^2$ initially oscillates due to the perturbation, grows slightly, and then remains mostly constant (or oscillates with a very low frequency). For the unstable case RHBH68, when the scalar field mass exceeds the black hole mass, $|\Phi|^2$ still exhibits a qualitatively similar behavior, but with larger oscillations persisting even after the initial transient. The total Noether charge, displayed in the bottom panels of Figure~\ref{fig:phi2max_Noether} and Figure~\ref{fig:phi2max_Noether_norm}, is not significantly affected by the dynamics of either the stable or unstable RHBH and remains constant throughout the entire simulation with very high accuracy. 

\begin{figure}
	\centering
	\includegraphics[width=0.45\textwidth]{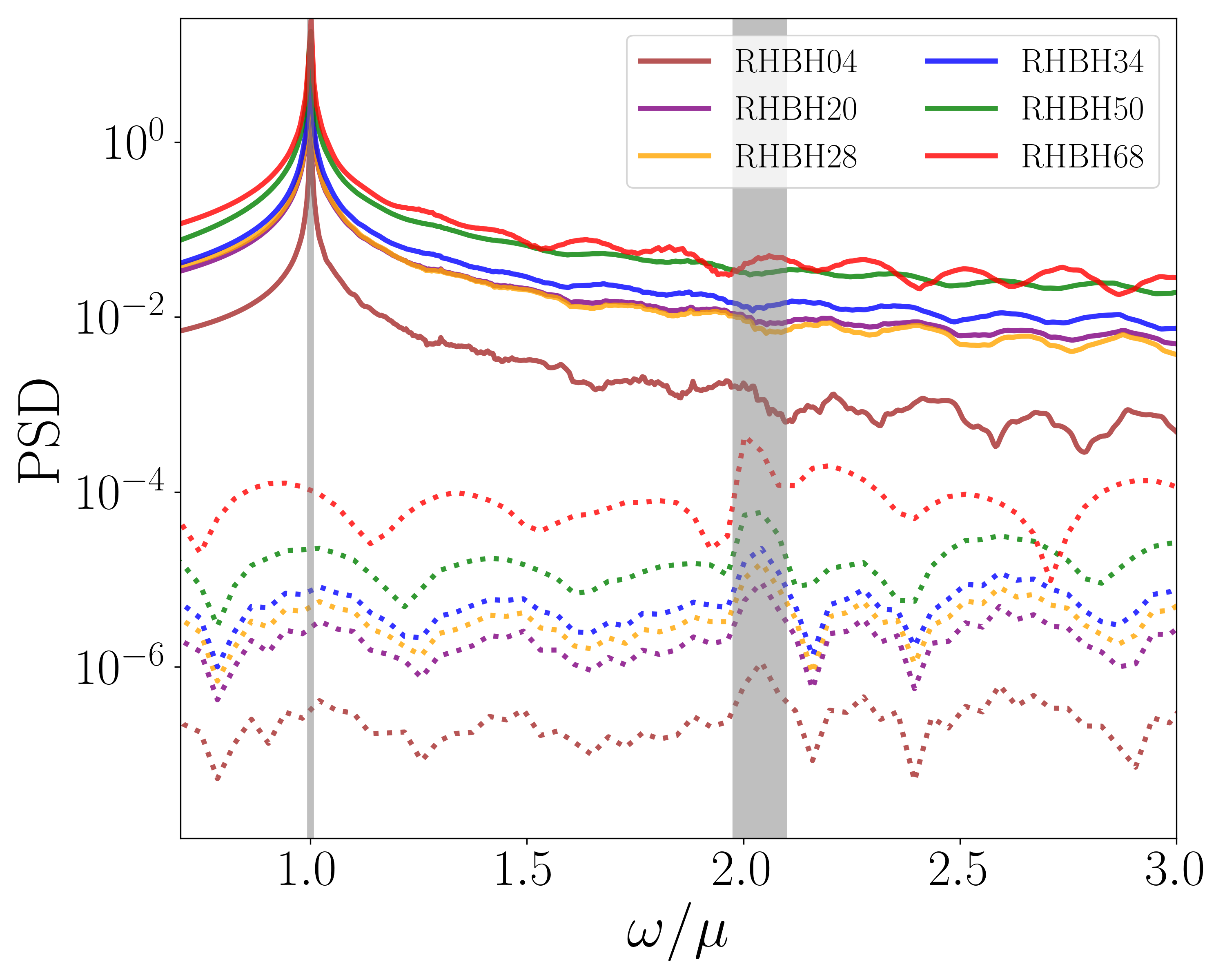} 
	\caption{{\em Dynamics of the RHBH configurations}. Fourier transforms of two quantities are shown: the real part of the scalar field, $\Phi_R$ (solid lines), and of its density, $|\Phi|^2$ (dashed lines). The spectrum of $\Phi_R$ exhibits a dominant peak at $\omega/\mu \approx 1$, as expected from the synchronization condition $\omega = \Omega_{\rm BH}$ (recall that $\mu \approx \Omega_{\rm BH}$). In contrast, the oscillations in  $|\Phi|^2$ exhibit a much richer spectral structure. The dominant peak appears at a higher frequency, approximately ${\bar \omega} \simeq 2 \Omega_{\rm BH}$, which could induce analogous perturbations in the spacetime metric.
	}    
	\label{fig:Fourier_spectrum} 
\end{figure}

Finally, we turn our attention to the analysis of the scalar field oscillation frequencies. The oscillations in the real (or imaginary) component of the field arise from its natural harmonic dependence and are expected to match the black hole’s angular velocity $\Omega_{\rm BH}$ according to the synchronization condition Eq.~(\ref{synchronization}). In contrast, the oscillations in $|\Phi|^2$ correspond to modes of the matter field in the torus, which have not yet been explored through perturbative analyses. To compare these behaviors, we plot in Figure~\ref{fig:Fourier_spectrum} the Fourier spectra of both quantities. As expected, we find that the oscillation frequency satisfies $\omega \approx \mu \approx \Omega_{\rm BH}$. More interestingly, the oscillations of the scalar field torus occur at approximately twice this frequency, ${\tilde \omega} \approx 2 \Omega_{\rm BH}$, regardless of the scalar field mass fraction in the RHBH configuration. A more detailed perturbative analysis will be required to fully understand this behavior.

\subsection{Stability analysis}
\label{sec:stability_analysis}

\begin{figure}
	\label{fig:snap}
	\centering
	\
	\includegraphics[width=1.\linewidth]{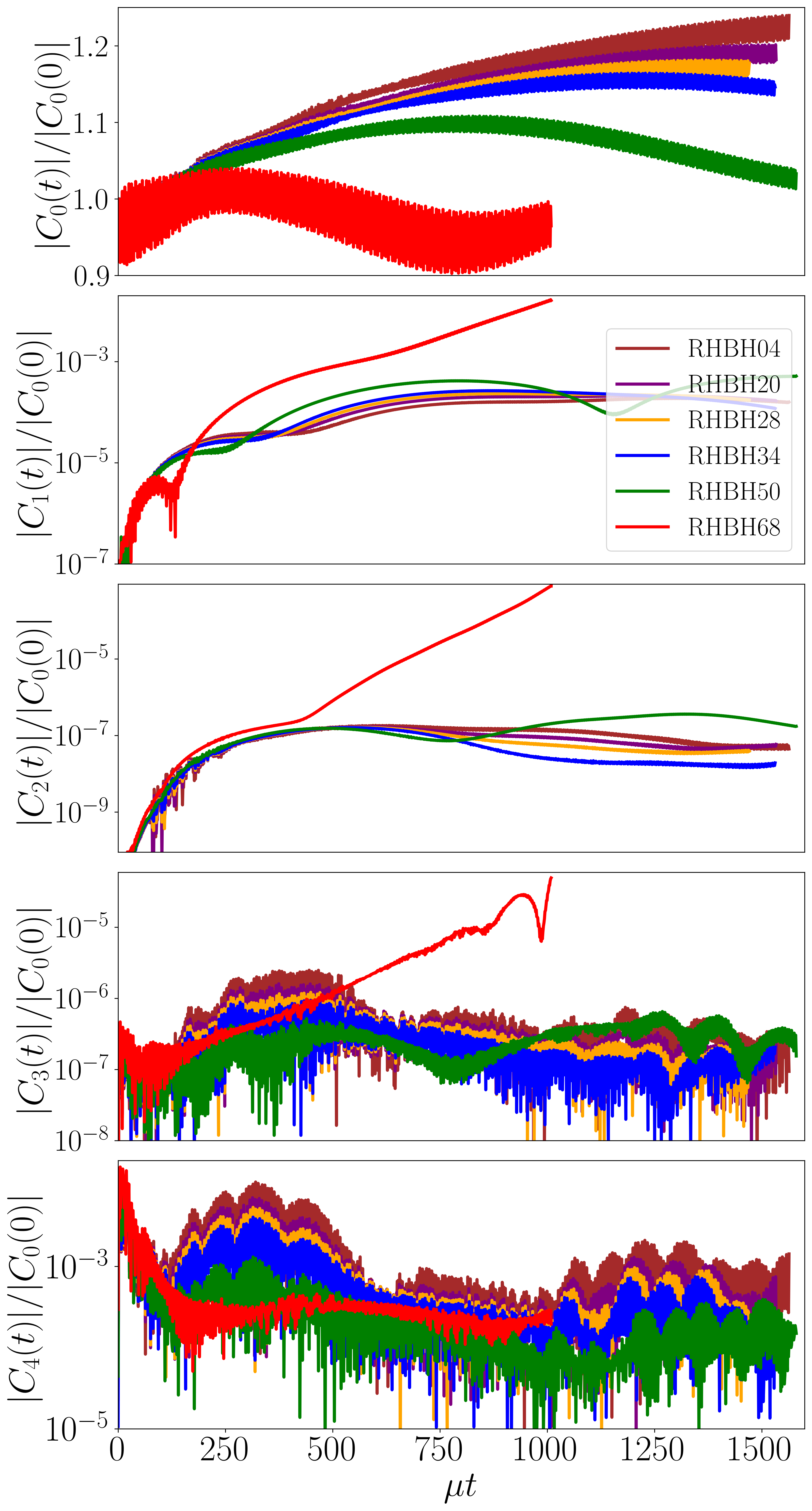}
	\caption{\label{fig:modes} {\em Dynamics of the RHBH configurations}.
          The plots represent the decomposition of \( |\Phi|^2 \) given by Eqs. (\ref{eq:azimuthal_decomposition},\ref{eq:azimuthal_decomposition2}). For all cases with $M_{\Phi}/M \lesssim 0.5$, the behavior follows the same qualitative trend: an initial small growth followed by a stabilization phase, both dominated by the $m=0$ mode. The only exception is the $m=4$ mode, which is strongly excited at early times by the initial perturbation, followed by a slow decay. In contrast, the RHBH68 configuration exhibits a markedly different behavior. Although the system remains dominated by the $m=0$ mode throughout the evolution, the $m=1$ mode begins to grow exponentially around $t \approx 200$, subsequently triggering an exponential amplification of the higher modes $m=2$ and $m=3$. This cascade leads to a significant redistribution of the system’s mass. }
\end{figure}

The previous qualitative and quantitative analyses suggest that RHBHs with $M_{\Phi}/M \lesssim 0.5$ are stable, at least for moderate spins $J^2_{\rm BH}/M_{\rm BH}=0.5$. To gain a deeper understanding of the onset of instability, we perform a more detailed analysis of $|\Phi|^2$ in the $z=0$ plane. In this plane, the scalar field density surrounds the black hole, forming an axisymmetric ring. We therefore perform an azimuthal decomposition of this quantity at constant radius $r=R$ and polar angle $\theta=\pi/2$, namely
\begin{equation}
	|\Phi|^2 = \sum_{m=0}^{\infty} b_m(R,t)\, e^{i m \varphi},
	\label{eq:azimuthal_decomposition}
\end{equation}
where the coefficients $b_m(R,t)$, computed numerically by integrating $|\Phi|^2$ over annular regions within a selected radial range, quantify the contribution of each azimuthal mode $m$ at a given radius $R$. A more compact representation of these modes can be obtained by performing a radial integration of these coefficients,
\begin{equation}
	C_m(t) = \int_{R_{\rm BH}}^{R_{\rm max}} b_m(R,t) \, R dR,
	\label{eq:azimuthal_decomposition2}
\end{equation}
where $R_{\rm BH}$ denotes the radius of the BH horizon and $R_{\rm max}$ provides an estimate of the outer radius of the scalar field torus. The modulus $|C_m(t)|$ provides a direct measure of the strength of each mode, allowing us to identify the global dominant azimuthal structures and to track their temporal evolution. Monitoring these modes over the full time series enables a quantitative characterization of the instabilities and their growth dynamics.

Figure~\ref{fig:modes} shows the relative amplitude of the azimuthal modes up to $m=4$, rescaled by the initial value of $C_0(t=0)$, for each RHBH configuration. In all cases, the dominant contribution corresponds to the $m=0$, which oscillates with two characteristic frequencies after the initial transient: a fast component, previously discussed, ${\tilde \omega}_f \approx 2 \Omega_{\rm BH}$, and a much slower one ${\tilde \omega}_s \approx 10^{-3} \Omega_{\rm BH}$. In both cases, the amplitude remains bounded throughout the evolution and close to its initial value. 
The higher-order modes are rapidly excited by the initial perturbation; however, they either remain bounded or decay over time. In contrast, for the RHBH68 configuration, all modes with $1 \leq m < 4$ exhibit exponential growth, indicating a breakdown of axisymmetry. This behavior is particularly pronounced in the $m=1$ mode, which acts as the primary driver of the instability in the whole system.

The behavior of this instability is similar to the non-axisymmetric instability (NAI) already observed in numerical simulations of rotating (scalar) boson stars with purely massive potentials~\cite{NAI1,NAI2,NAI3}. In that case, the non-axisymmetric modes grow exponentially and ultimately disrupt the star. Given the similarity in behavior, we conjecture that the instability found in the RHBH corresponds to a NAI. This interpretation is consistent with the fact that, when the mass of the scalar-field torus dominates over the one of the black hole, the system effectively resembles a rotating boson star. One might be tempted to compare the instability timescales of our RHBH configurations with the NAI timescales of boson stars reported in the literature (see, for instance, Figure 6 in Ref.~\cite{NAI2}). For a broad range of boson star compactness, these timescales lie in the interval $T/M \sim {\cal O}(10)-{\cal O}(10^3)$, with more compact configurations exhibiting shorter instability times.
However, a direct comparison with our RHBH configurations is not straightforward, since the presence of the black hole significantly alters both the compactness and the structure of the bosonic cloud, except in the limit $M_{BH}/M \rightarrow 0$. The closest configuration we were able to construct near this limit is RHBH68, which has an angular momentum $J/M^2 \sim 1.5$. A boson star with comparable angular momentum would be expected to exhibit an instability timescale $T_{NAI}/M \sim {\cal O}(10^3)$, corresponding to $\mu T_{NAI} \sim {\cal O}(10^2)$. This is consistent, at the order-of-magnitude level, with the timescale observed for the development of the instability in RHBH68.
We also note that, in the unstable RHBH case, a growing mode can be identified relatively quickly. By contrast, for the remaining configurations we observe no sign of instability over the duration of our simulations, corresponding to $T_{NAI}/M \gtrsim {\cal O} (10^4)$, i.e. more than an order of magnitude longer than the longest NAI timescales reported for rotating boson stars. Furthermore, in \cite{NAI1} the authors find that 
for initial stationary rotating boson star configurations, which again do not necessarily compare with RHBH68, but that are also in the fundamental state (no nodes), the NAI starts developing
around $t\mu= 10^3$, where the star fragments in two and then the binary system remains bounded gravitationally for a while and then merges into a BH around $t\mu= 1.2\times 10^3$. So the scale of instability is also of the same order of magnitude found in RHBH68. In order to gain a much more rigorous correlation between the non-linear instability found for configurations similar to RHBH68 (i.e. configurations where $M_{\Phi}/M> 0.5$) and the NAI, a linear stability analysis is needed, like the one performed in \cite{Ganchev2018} but having as a background exactly the
same initial unperturbed configuration that is evolved in the non-linear case. Only then one might see
if the time scale  $\omega_I T\sim 1$ found in the linear analysis associated with the
mode of frequency $\omega_I$ matches the scale $\mu T$ found in the non-linear
evolution. A systematic study along these lines together with
a parallel analysis of comparable configurations of scalar rotating boson stars
can shed light on the possible correlations between the NAI of both
kind of objects.

\begin{figure}
	\label{fig:snap}
	\centering
	\includegraphics[width=0.8\linewidth]{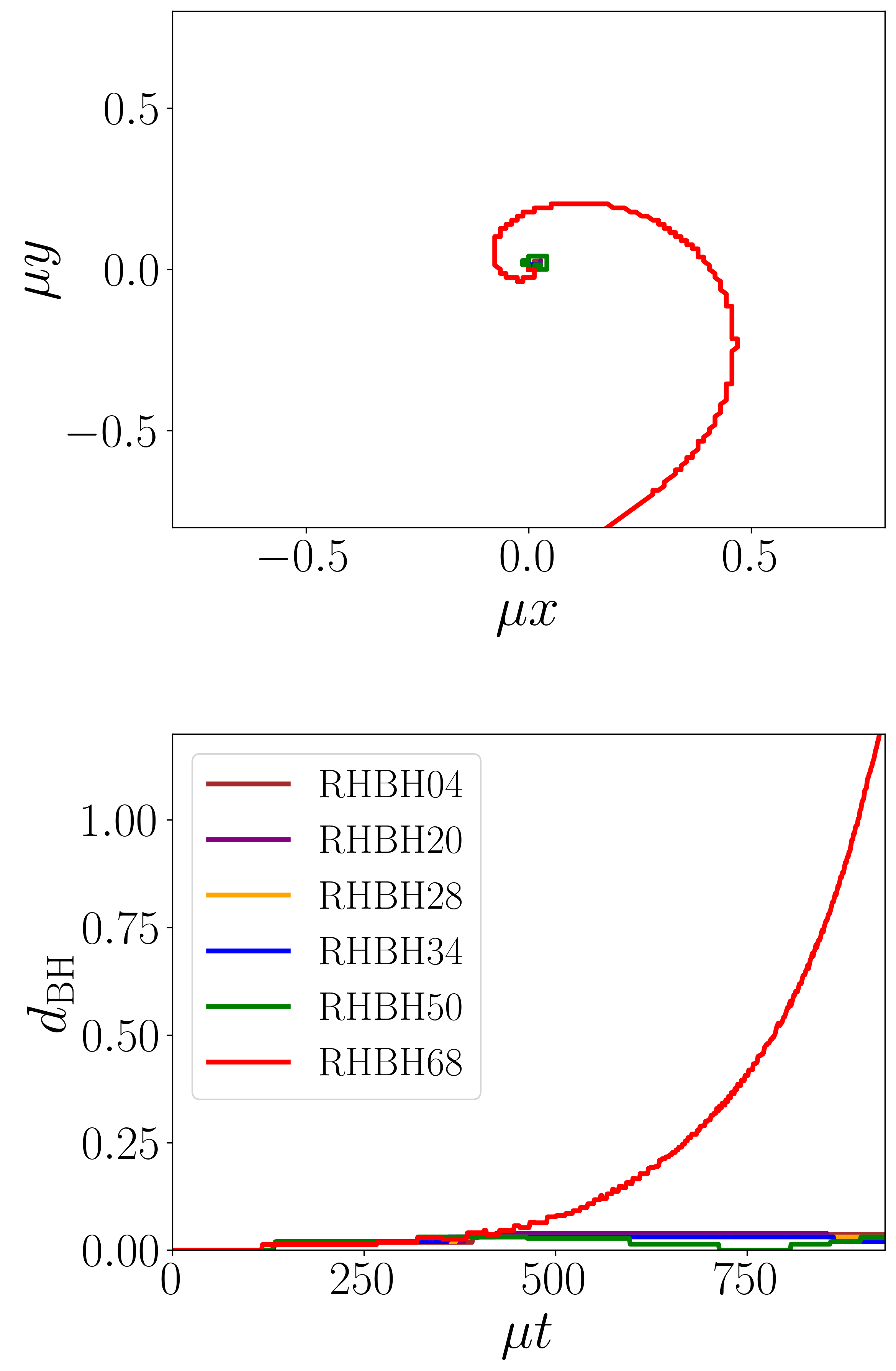}
	\caption{\label{fig:BHdance} {\em Dynamics of the RHBH configurations}. The position of the BH, computed as the location of the minimum of the lapse, is shown in the $x$–$y$ plane (top) and as the cylindrical radius of this displacement (bottom). In all configurations analyzed, except for RHBH68, the black hole remains essentially at the origin of the coordinate system. In the few cases where a slight displacement is observed, it is purely oscillatory around the equilibrium position, indicating a stable dynamics. The RHBH68 configuration exhibits a quite different behavior: the distance of the black hole from the origin grows exponentially, resulting in a spiral trajectory. Ultimately, this motion drives the black hole to collide with the surrounding torus, leading to its disruption.}
\end{figure}

Another consequence of this instability is that the black hole drifts away from the origin, following an outward spiraling trajectory that gradually increases in radius until it collides with and disrupts the torus. The black hole trajectories, computed by tracking the minimum of the lapse function $\alpha$, are shown in Figure~\ref{fig:BHdance}. In contrast, for the stable configurations, the initial position of the black hole corresponds to a stable equilibrium point. During the evolution, it only oscillates slightly around its initial position.
As discussed in \cite{2025arXiv250920450N}, this behavior can be understood with a simple Newtonian model of a test particle moving within a ring of matter distribution. Although this model provides a good approximation when most of the mass resides in the scalar-field torus, it fails to accurately describe the opposite regime. Here we have extended such model as follows: the scalar field torus is represented by a ring of constant density matter, while the black hole is modeled as a uniform density sphere. By analyzing the stability of a test particle near the origin, we identify two competing contributions. The ring induces a term $\sim -r^2$ in the gravitational potential, whereas the sphere contributes with a term proportional to $\sim r^2$. The total gravitational potential near the center has the form
\begin{equation}
	V(r) \approx \text{constant} + r^2 \left( \frac{M_s}{2R_s^3} - \frac{M_r}{4R_r^3} \right).
\end{equation}	
where $M_s$ is the mass of the sphere with radius $R_s$, and $M_r$ is the mass of the ring with radius $R_r$. The sign of the coefficient of $r^2$ determines the nature of the equilibrium point near the origin. Consequently, when the gravitational influence of the black hole dominates (i.e., when the coefficient of $r^2$ is positive), perturbations are damped and decay over time. In contrast, when the gravitational contribution of the scalar-field torus becomes dominant, the coefficient of $r^2$ turns negative, and perturbations grow exponentially, signaling an unstable equilibrium. 

\section{Conclusions}
\label{sec:discussion}

The stability of RHBHs has been one of the central questions since their first construction in the seminal works~\cite{Herdeiro2014,Herdeiro2015}. Some insight into their stability properties can be gained by analyzing limiting cases. 

Near the test-field limit, $M_{\Phi}/M \rightarrow  0$,  RHBH configurations (i.e., the so-called clouds) were studied in~\cite{Ganchev2018}, where they were found to be unstable and to resemble weakly perturbed Kerr solutions. It remains unclear whether the instability reported in~\cite{Ganchev2018} is a transient effect that eventually settles into an RHBH with a larger scalar-field mass contribution, or whether these configurations are marginally unstable of the kind referred to as effectively stable (that is, systems whose instability timescale exceeds or
are close to the age of the Universe)~\cite{Degollado2018}. Indeed, the instability in those cases develops only over timescales of order 
$t\mu \sim {\cal O} (10^{11})$, making its ultimate outcome difficult to predict. In any case, such extremely long timescales are far beyond those accessible to the fully nonlinear numerical evolutions like the ones performed in this work.

On the other limit, $M_{\Phi}/M \rightarrow  1$, the RHBH configurations approach rotating boson stars containing a tiny central black hole. Such configurations, for a purely massive potential, have been shown to suffer from a linear non-axisymmetric instability (or NAI), which typically develops on short timescales. Therefore, it remained unclear for which fraction of $M_{\Phi}/M$ the system becomes unstable, or whether the transition between stable and unstable regimes is instead continuous.

The fully nonlinear simulations of the Einstein-Klein-Gordon system presented here strongly suggest that equilibrium RHBH configurations are stable, at least for moderate spins  $J_{\rm BH}/M^2_{\rm BH} \lesssim 0.5$,
whenever the black hole mass dominates over that of the scalar field torus, i.e., for 
$M_{\Phi}/M \lesssim 0.5$. Within this regime, no signs of superradiant or non-axisymmetric instabilities are observed in timescales of $t\mu \sim {\cal O} (10^{3})$. Therefore, RHBHs with a subdominant scalar-field contribution and moderate spin appear to be {\it effectively} stable and thus astrophysically viable. Notice also that these results are consistent with the recent simulations presented in~\cite{2025arXiv250920450N}.

It is also worth commenting on other related hairy black hole (BH) solutions, such as those arising in the Einstein-Proca system, which involve a complex, massive vector field (Proca hair). In Ref.~\cite{East-Pretorius2017}, the authors showed that, due to superradiant instabilities, Proca hair can form around a nearly extremal Kerr black hole with $J_0/M_0^2=0.99$, given in terms of the initial mass $M_0$ and initial angular momentum $J_0$ of the Kerr BH. In this process, approximately between 2–9\% of the initial mass and 5–38\% of the initial angular momentum can be extracted, depending on the range of $M_0 \mu \sim [1/4,1/2]$. For reference, an extremal Kerr BH can in principle radiate up to 
29\% of its mass, corresponding to the difference between its total mass and its irreducible mass~\cite{Wald1984}. Notwithstanding, in Ref.~\cite{Herdeiro2021} the authors
argue that if the superradiant evolution is approximately
{\it conservative} then only up to 10\% of its mass could be effectively radiated, regardless
of the spin of the boson hair. The timescale for the development of the instability and the subsequent formation of a Proca hairy BH is of order $t\mu \sim {\cal O} (10^{4})$, with only modest dependence on $M_0$. This corresponds to the time required for the Proca field frequency $\omega_P$ to synchronize with the BH horizon frequency (for $m=1$). Therefore, some of the stationary Proca RHBH solutions constructed in Ref.~\cite{Herdeiro2016} are likely to represent the final states of the superradiant instability of Kerr BHs, as observed in the fully nonlinear evolutions of Ref.~\cite{East-Pretorius2017}. 
In contrast, in the scalar (boson) case discussed here, we do not observe superradiant instabilities within comparable timescales. Linear perturbation estimates indicate that the growth times of these instabilities are up to seven orders of magnitude longer, and in the most optimistic cases, four orders longer than in the vector boson scenario~\cite{Dolan2007}, corresponding to 
$\mu t \sim {\cal O} (10^{11})$ or $\mu t \sim {\cal O} (10^{7})$, respectively. 

Nevertheless, although we cannot numerically model the formation of RHBHs composed of scalar fields through the superradiant instability (i.e., due to the extremely long associated timescales), we can reasonably assume that they might form through a similar mechanism, allowing us to draw some conclusions about their final configurations.
If $M_0$ denotes the initial ADM mass of the Kerr black hole and $M_{\rm BH}$ the final black hole mass after the hair has formed, then in principle
$M_{\Phi} = M_0-M_{\rm BH} < M_0-M_{\rm irr}$.
For an extremal Kerr black hole, where the radiated energy is maximal, this yields $M_{\Phi} < M_0(1-M_{\rm irr}/M_0) \lesssim 0.29 M_0$. As previously discussed, RHBH configurations with $M_{\Phi}/M \lesssim 0.5$ are found to be
{\it effectively} stable, regardless of the mechanism responsible for their formation. We may therefore conclude that all RHBHs formed through the superradiant instability are expected to be stable against non-axisymmetric and other types of instabilities, at least over moderate timescales. It is, however, possible that some RHBHs are produced through mechanisms other than superradiance; in such cases, the emergence of unstable configurations cannot be ruled out.

The conclusions drawn in this work regarding the non-linear stability of rotating hairy black holes rely on solid numerical evidence, further reinforced by physical intuition gained from the well-established behavior of Kerr black holes, other families of hairy black holes, and rotating boson stars.
Our simulations strongly indicate the existence of an
{\it effectively} stable branch of RHBH configurations, particularly when the scalar field contribution remains moderate with respect to the total one. Notwithstanding, a complete understanding of their stability properties demands a systematic exploration through linear perturbation analyses around the equilibrium configurations employed here as initial data, complemented by longer and higher-resolution non-linear evolutions.
Such an extended study, aimed at mapping precisely the boundaries between stable and unstable regimes in the RHBH parameter space, will be pursued in future works.


\section*{Acknowledgments}
This work was supported by the project PID2022-138963NB-I00, funded by the Spanish Ministry of Science, Innovation and Universities (MCIN/AEI/10.13039/501100011033). JAC acknowledges support from a predoctoral fellowship (FPI) associated with this project (reference PREP2022-000480).
This work was partially supported by 
DGAPA-UNAM grant IN105223. M.S. acknowledges support from DGAPA-PASPA sabbatical grant.
The authors thankfully acknowledges the computer resources at MareNostrum and the technical support provided by Barcelona Supercomputing Center (RES-AECT-2025-2-0007 and RES-AECT-2025-2-0024). This work was granted access to the HPC resources of MesoPSL financed by the Region Ile de France and the project Equip@Meso (Reference No. ANR-10-EQPX-29-01) of the programme Investissements d’Avenir supervised by the Agence Nationale pour la Recherche.


\appendix

\section{Reliability of the numerical solutions}
\label{appendixA}

Our initial data correspond to the RHBH solutions presented in Section~\ref{sec:numerical_ID}, with an additional strong perturbation introduced by artificially increasing the numerical discretization errors at the boundaries between refinement levels (i.e., by using low-order spatial discretization operators which are second-order accurate on the faces and zeroth order at corners and edges). This procedure introduces errors, in the form of several cubic shells, both in the evolution equations and in the constraints. These perturbations propagate until the system relaxes to a stationary configuration. Because they break axial symmetry, they excite a broad range of modes. Although most of such errors are expected to vanish in the limit of infinite resolution, they are relatively large for the resolutions employed in this work. It is therefore essential to verify the reliability of the numerically obtained solutions by demonstrating convergence to the continuum limit and ensuring that our results are robust under changes in the initial perturbation.

Although the nature of our perturbation is convenient to excite a broad range of modes, it is not well suited for convergence tests, since the perturbation in the scalar field depends on the resolution in a non-trivial way. Furthermore, the simulations are computationally very expensive, which prevents a systematic exploration of different cases as well as their extension to longer time intervals. Therefore, we do not aim to demonstrate strict convergence rates, which have nonetheless been established for our code, using the same evolution system and numerical techniques, in the context of boson stars (see Refs.~\cite{2017PhRvD..95l4005B,2017PhRvD..96j4058P,2022PhRvD.105f4067B}). Instead, we provide evidence of qualitative convergence and numerical robustness to support the physical conclusions of this work by presenting results at three different resolutions. The lowest resolution is the one shown throughout the paper and corresponds to the minimum resolution at which we are confident that both the black hole and the boson cloud are sufficiently well resolved. The grid spacing is reduced by a factor of 1.25 in the medium-resolution simulation, and by the same factor again in the high-resolution one. We monitor global quantities, such as the constraints and the Noether charge, to ensure that fundamental conservation laws remain satisfied within acceptable numerical errors. In addition, we track the evolution of key variables relative to their initial values in order to identify potential numerical instabilities.

\begin{figure}[H]
	\centering
	\includegraphics[width=1.0\linewidth]{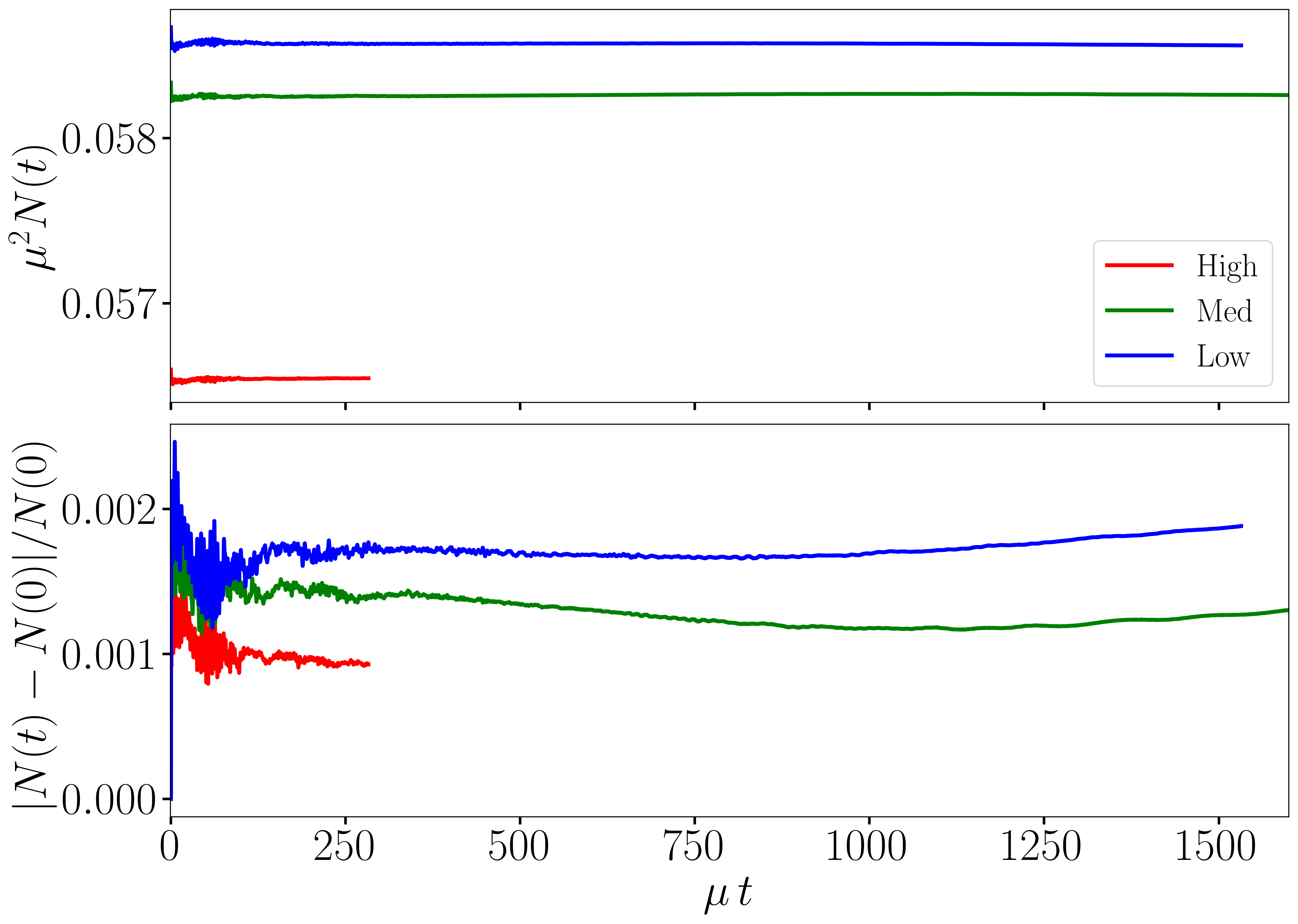}
	\caption{\label{fig:des_noef} {\em Resolution study of the configuration RHBH34.} Evolution of the Noether charge and its relative deviation from the initial value for different resolutions in configuration \textbf{RHBH34}. The initial value varies with resolution due to the increased grid accuracy and the nature of our perturbation. The normalized deviation highlights the excellent conservation throughout the simulation, with errors below 
	$0.2\%$  and convergence toward zero. This behavior supports the global consistency of the simulation. }
\end{figure}

In Fig.\ref{fig:des_noef} we show the total Noether charge 
$N(t)$ (top panel) together with its rescaled quantity 
$|N(t)-N(0)|/N(0)$ (bottom panel), which represents the relative deviation from its initial value. For the three resolutions considered, the scalar charge remains nearly constant: the deviation from the initial value is small and, moreover, all resolutions exhibit the same behavior as the simulation progresses. This indicates that the system’s global dynamics converge correctly, although the initial perturbation slightly modifies the initial value in a convergent manner.

\begin{figure}
	\centering
	\includegraphics[width=1.0\linewidth]{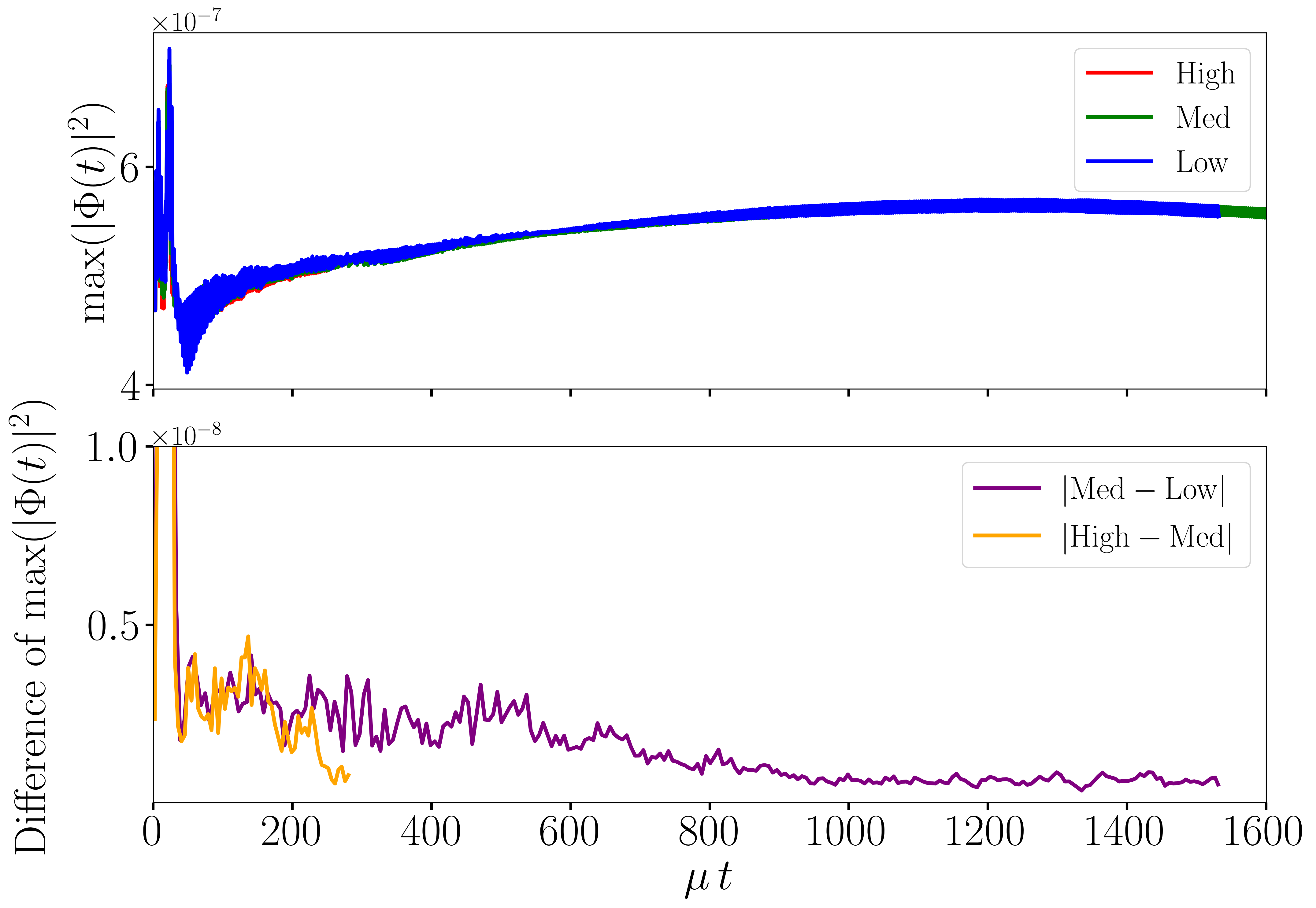}
	\caption{\label{fig:nor_phi2} {\em Resolution study of the RHBH34 configuration.} Evolution of the maximum of $|\phi|^2$ and the difference of this value between resolutions. The maximum follows the same behavior for the three resolutions studied. In the bottom panel it can be seen that, when the system relaxes for $\mu t>200$, the difference between medium and high resolutions is roughly by a factor 2 than between medium and low resolutions, indicating a convergence close to third order.}
\end{figure}

Another relevant quantity is the maximum of $|\phi|^2$, displayed in the top panel of Fig. \ref{fig:nor_phi2}, along with the absolute difference between resolutions shown in the bottom panel. For the short initial transient at 
$\mu t<100$, the nature of the perturbation (i.e., being zeroth order at corners and edges) prevents this quantity (which corresponds to a maximum rather than an integrated measure) from exhibiting the expected convergence behavior. The longer oscillation around $\mu t \sim {\cal O} (1000)$ arises from a combination of two effects: (i) a deviation from the initial stationary solution induced by the perturbation, and (ii) the dynamical gauge evolution, which drives the system away from the coordinate frame in which the solution was originally stationary.
Nevertheless, all three resolutions exhibit essentially the same behavior during the time evolution, indicating that, for the resolutions considered, the results are largely independent of the grid discretization. As shown in the bottom panel, after the system has relaxed $\mu t>200$, the difference between the medium and high-resolution results is approximately a factor of 2 smaller than that between the medium and low-resolution ones, roughly consistent with third order convergence following the initial transient. The weak dependence on spatial resolution and the consistent temporal behavior across all resolutions, both during the initial transient and in the subsequent quasi-stationary phase, suggest that the dynamical response of the scalar field to the perturbation is robust.

\begin{figure}[H]
	\centering
	\includegraphics[width=1.0\linewidth]{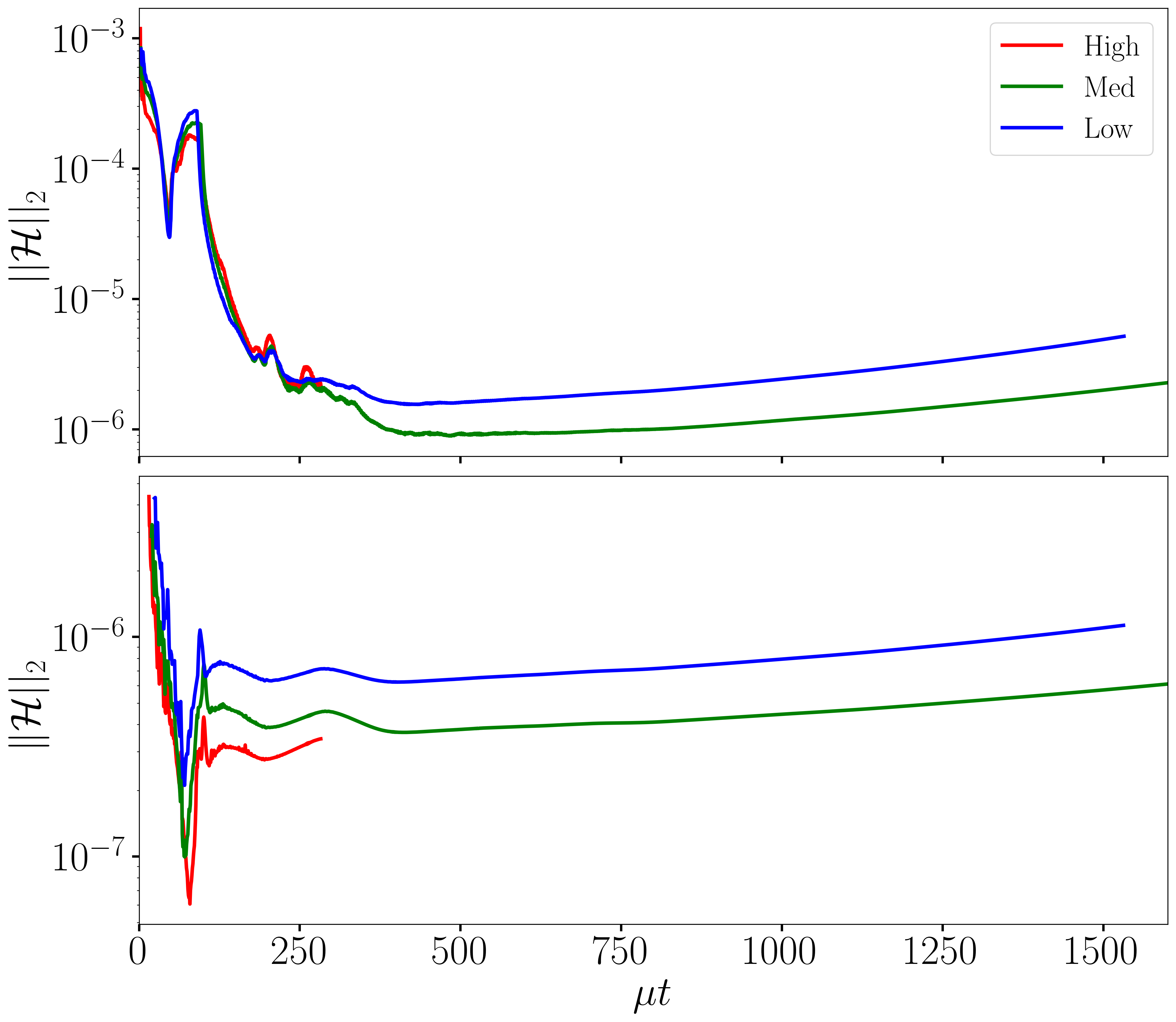}
	\caption{\label{fig:ham_con} {\em Resolution study of the configuration RHBH34.} Evolution of the Hamiltonian constraint for different resolutions in configuration \textbf{RHBH34}. The top panel shows the 
	$L^2$-norm over the full domain, while the bottom panel shows the same quantity restricted to the torus containing the scalar-field cloud. The constraint-damping terms enforce the constraints in both regions, with a more pronounced effect in the black-hole interior, where the initial data do not satisfy the constraints. Increasing the resolution leads to convergence toward zero in both regions, more clearly in the domain outside the black hole horizon.}
\end{figure}

The $L^2$-norm of the Hamiltonian constraint is shown in Fig. \ref{fig:ham_con}, both over the entire computational domain (top) and restricted to the exterior of the apparent horizon (bottom). Since the interior of the black hole is constructed via interpolation of the initial data, there are large constraint violations inside the event horizon at the initial time which are not expected to converge away with increasing resolution. As the evolution proceeds, the constraint damping associated to the Z4 formulation suppresses most of these violations, reducing the constraint by several orders of magnitude. After this transient phase, the Hamiltonian constraint remains well controlled throughout the simulation, exhibiting only a small growth at late times. We emphasize that a more carefully tuned choice of damping parameters could potentially slow down or even eliminate this late-time growth.

{}

\end{document}